\documentclass[prd,onecolumn,preprintnumbers,floats,floatfix,apsrev,amsmath,amsfonts,nofootinbib,superscriptaddress]{revtex4-2}
\usepackage{color}
\usepackage{graphicx}
\usepackage{dcolumn}
\usepackage{natbib}
\usepackage{hyperref}
\usepackage{hypernat}
\usepackage{natbib}
\usepackage{soul}
\usepackage{amssymb}

\newcommand{\dd}{{\rm d}}

\newcommand{\p}{\partial}

\newcommand{\be}{\begin{equation}}
\newcommand{\en}{\end{equation}}

\newcommand{\paren}[1]{\left({#1}\right)}

\newcommand{\ex}{\mathrm{ex}}

\begin{document}
\title{Thermodynamical and dynamical stability of Einstein-Maxwell and extremal Einstein-Born-Infeld thin shells in $(2\ \mathbf{+}\ 1)$ dimensions}

\author{Dario Olmos Cayo}
\email{dolmosc@fcpn.edu.bo}
\author{Zui Oporto Almaraz}
\email{zoporto@fiumsa.edu.bo}
\affiliation{Instituto de Investigaciones Físicas, and Planetario Max Schreier, Universidad Mayor de San Andrés, Campus Universitario, C. 27 s/n Cota-Cota, 0000 La Paz, Bolivia.}

\author{M.~L.~Pe\~{n}afiel}
\email{miguelpenafiel@upb.edu}
\affiliation{UERJ - Universidade do Estado do Rio de Janeiro 150, CEP 20550-013, Rio de Janeiro, RJ, Brazil.}
\affiliation{FIA - Facultad de Ingeniería y Arquitectura, Universidad Privada Boliviana, Camino Achocalla Km 3.5, La Paz, Bolivia.}

\date{\today}

\begin{abstract}
 We study the dynamical and thermodynamical stability of thin shells in (2+1)-dimensional spacetimes composed of an inner anti-de Sitter (AdS) region and an outer region described by a charged Bañados–Teitelboim–Zanelli (BTZ) spacetime, sourced either by Einstein–Maxwell theory (Maxwell-BTZ) or Einstein–Born–Infeld theory (BI-BTZ). Assuming a fixed charge-to-mass ratio and modeling the shell's matter with a linear equation of state, we introduce a convenient parametrization to analyze the dynamical stability configurations. We find that Maxwell-BTZ thin shells admit a wider range of dynamically stable configurations compared to BI-BTZ thin shells. We also derive the thermodynamics of the shell matter, obtaining physically meaningful entropy functions in both cases, and examine the conditions for thermodynamical stability. In the Maxwell-BTZ case, we identify regions in the parameter space where configurations are both dynamically and thermodynamically stable. In contrast, for extremal BI-BTZ thin shells, all thermodynamically stable configurations are contained within the dynamically stable ones, and shells with a linear equation of state are always dynamically stable. This work extends the understanding of thin shell configurations in lower-dimensional gravity and elucidates the interplay between dynamics, thermodynamics, and nonlinear electrodynamics.
\end{abstract}

\maketitle
\section{Introduction}

The study of gravitational models in $\paren{2+1}$ dimensions offers some unique insights for the physics underlying gravity that are often difficult to obtain when studying gravity in higher dimensions. Despite the lack of local degrees of freedom, gravity in $\paren{2+1}$ dimensions admits a rich set of exact, nontrivial solutions including black holes \cite{Banados1992,Banados1993,Carlip1995,Martinez2000}, conical defects \cite{Deser1984} and wormholes \cite{Perry1992}. Also, it is worth emphasizing that $\paren{2+1}$-dimensional spacetimes may permit a canonical quantization of gravity, as long as there exists a non-trivial cosmological constant \cite{Achucarro1986,Witten1988,Constantinidis2015}. Moreover, several approaches for obtaining a quantum theory of gravity can be implemented in a $\paren{2+1}$-dimensional spacetime (see Refs.~\cite{Carlip2005,Carlip2023} and references therein). 

The $\paren{2+1}$-dimensional black hole arises as the endpoint of gravitational collapse of matter or radiation \cite{Ross1993,Husain1994}, following an analog of the Oppenheimer–Snyder scenario, in which a thin shell of matter collapses to form a black hole. These black holes also exhibit intriguing thermodynamic properties and provide connections to quantum gravity, further motivating their study.

Thin shells offer an idealized but powerful framework to model gravitational collapse and matter distributions in general relativity. These configurations are constructed by gluing two spacetime regions across a hypersurface of matter, with the junction conditions ensuring that the resulting solution is well-defined, as introduced by Israel \cite{Israel1966}.
A particularly interesting class of thin shell spacetimes consists of an inner region devoid of matter and an outer region described by a nontrivial geometry (such as a black hole) joined continuously at the shell. Thin shell solutions have also been explored within general relativity but with alternative theories for the matter sector, such as nonlinear electrodynamics \cite{Eiroa2012}, and in the context of extended theories of gravity \cite{Eiroa2018}.

Theories of nonlinear electrodynamics (NLED) have been proposed as generalizations of Maxwell’s theory to describe phenomena beyond the reach of the linear theory. Notably, Born–Infeld (BI) electrodynamics \cite{Born1934} was originally developed to regularize the electric field and energy of a point charge, and it later gained prominence due to its appearance in the low-energy limit of string theory \cite{Fradkin1985}. When coupled to gravity, NLED models can yield regular black hole solutions \cite{AyonBeato1998,Dymnikova2004}, generalize black hole thermodynamics \cite{Gulin2017,Bokulic2021}, and affect light propagation in black hole spacetimes \cite{Guzman‐Herrera2022,GuzmanHerrera2024}, among other features.

In the context of $\paren{2+1}$ dimensions, the black hole solution within BI electrodynamics — often referred to as the Born-Infeld-BTZ (BI-BTZ) black hole — was first derived by Cataldo and Garcia \cite{Cataldo1999}. Its thermodynamics was subsequently studied in \cite{Myung2008}, and its stability within the canonical ensemble was investigated in \cite{Hendi2012}.

The stability of thin shells can be assessed from two complementary perspectives. On the one hand, one can analyze the \emph{dynamical stability} by perturbing the shell's equation of motion and examining its response to radial perturbations. On the other hand, one can examine the \emph{thermodynamical stability} by requiring that internal changes of energy within the shell do not lead to a phase transition, which translates into conditions on the derivatives of the shell's entropy \cite{Martinez1996}.

These two criteria are conceptually different and independent; however, one may wonder whether a thin shell solution that fulfills both stability criteria simultaneously can be regarded as more stable than those satisfying only one of them \cite{Bergliaffa2020,Reyes2022,Eiroa2024}. 

In the context of $\paren{2+1}$ dimensions, the thermodynamics and thermodynamical stability of various thin shell configurations have been studied in \cite{Lemos2013,Lemos2014,Lemos2015a}, while the construction and linear stability of BI thin shell was addressed in \cite{Eiroa2013}.

This paper studies both dynamical and thermodynamical stability criteria for a $\paren{2+1}$-dimensional thin shell composed of an AdS inner region and a BI-BTZ outer region, and contrasts the results with the corresponding Maxwell limit.  In Section~\ref{sec:BIBTZ} we review the BI-BTZ black hole solution, ensuring that the Maxwell-BTZ and BTZ limits are properly recovered. In Section~\ref{sec:TSST} we construct the thin-shell spacetime by matching an inner AdS region to an outer BI-BTZ region across a matter shell. In Section~\ref{sec:dynstab} we study the dynamical stability of the shell: we independently analyze the BI theory and its Maxwell limit as separate cases. In Section~\ref{sec:THshell} we develop the thermodynamics of the matter composing the shell, beginning with the Maxwell case---which, to our knowledge, has not been previously explored---and then addressing the BI case, where we find that the extremal limit provides the most tractable scenario since in this particular case there exist a closed analytic expression for the gravitational radius in terms of the ADM mass and charge. Moreover, we explore two different equations of state for the inverse temperature that are well adapted for studying the cases under consideration: namely a Hawking-type and a power law temperature equations of state for treating the Maxwell-BTZ and extremal BI-BTZ cases respectively. Further, we derive analytic expressions for the entropy of these shells in an adequate parametrization that allows for the study of thermodynamical stability. In Section~\ref{sec:thermstab} we examine the thermodynamical stability of the Maxwell and extremal BI shells, and compare these results with the dynamically stable configurations. Finally, in Section~\ref{sec:conclusions} we summarize our findings and outlines perspectives for future work.

Throughout this work we set $G=c=\hbar=1$ and use the metric signature $\eta_{ab} = \text{diag}(+1,-1,-1)$. Lowercase Latin indices ($a,b,c$) denote $\paren{2+1}$-dimensional spacetime components $0,1,2$, and $i,j,k$ refer to the hypersurface components $0,1$.

\section{The Einstein-Born-Infeld black hole solution in $\paren{2+1}$ dimensions} \label{sec:BIBTZ}

In a $\paren{2+1}$-dimensional spacetime, Born-Infeld (BI) electrodynamics is a nonlinear generalization of Maxwell electrodynamics, that preserves gauge symmetry, for which the Lagrangian density is constructed as a function of the only scalar invariant constructed from the Faraday tensor, $F_{ab}=\p_a A_b-\p_b A_a$, namely\footnote{There exists another gauge invariant, up to a boundary term, the so-called Chern-Simons term $\Omega_{\text{CS}}=\epsilon_{abc}A^{a}F^{bc}$.}

\be
F\equiv\frac{1}{2}F^{ab}F_{ab}\ .
\en
The BI Lagrangian in $\paren{2+1}$ dimensions is given as \cite{Born1934,Cataldo1999}
\be
\label{eq:LBI}
\mathcal{L}_{\text{BI}}=\frac{\beta^2}{4\pi}\paren{1-\sqrt{1+\frac{F}{\beta^2}}}\ ,
\en
where $\beta$ is a maximum field parameter inherent to the theory. The linear theory is adequately recovered once the electromagnetic fields under assessment are taken as small compared to $\beta$.

We shall consider a static, circular metric of the form
\be
\dd s^2=f\paren{r}\dd t^2-f\paren{r}^{-1}\dd r^2-r^2\dd\phi^2\ ,
\en
which is a solution of the Einstein-Born-Infeld equations:
\begin{align}
R_{ab}-\frac{1}{2}g_{ab}R+\Lambda g_{ab}=8\pi T_{ab}\ , \\
T_{ab}=\frac{1}{4\pi}\left[-\frac{F_{ac}F^{c}_{\ b}}{\sqrt{1+F/\beta^2}}+\beta^2g_{ab}\paren{1-\sqrt{1+\frac{F}{\beta^2}}}\right] \ , \\
\nabla_aE^{ab}=0\ , \label{eq:ElecEq}
\end{align}
where 
\be
E^{ab}=\frac{F^{ab}}{\sqrt{1+F/\beta^2}}\ ,
\en
is the \emph{excitation} tensor. Notice that the electrostatic equation \eqref{eq:ElecEq} strongly resembles Maxwell's equations in material media. Since we are considering a static solution, we can take the {\em ansatz} $F_{ab}=E\paren{r}\paren{\delta_a^t\delta_b^r-\delta_a^r\delta_b^t}$ and the electrostatic equation can be written as
\be
\p_r\paren{\frac{r E\paren{r}}{\sqrt{1-E\paren{r}^2/\beta^2}}}=0\ .
\en
which has the solution
\be\label{eq:Efield}
E\paren{r}=\frac{Q}{\sqrt{r^2+Q^2/\beta^2}}\ ,
\en
where $Q$ is an integration constant identified as the electric charge. The electrostatic potential obtained from Eq. \eqref{eq:Efield} is
\be \label{eq:phiBIBTZ}
\phi=\frac{Q}{4} \ln\paren{\frac{\sqrt{1+\frac{Q^2}{r^2\beta^2}}-1}{\sqrt{1+\frac{Q^2}{r^2\beta^2}}+1}}\ .
\en
The nonvanishing components of the Einstein's tensor are
\begin{align}
G_{tt}&=\frac{f\paren{r}f'\paren{r}}{2r}\  , \label{eq:G00}\\
G_{rr}&=-\frac{f'\paren{r}}{2rf\paren{r}}\  , \\
G_{\phi\phi}&=-\frac{1}{2}r^2f''\paren{r}\ .
\end{align}	
The metric function $f\paren{r}$ can be obtained directly from plugging Eq. \eqref{eq:G00} into the Einstein's equation for the energy-momentum tensor corresponding to the field \eqref{eq:Efield}. The $\paren{2+1}$-dimensional black hole solution for the Einstein-Born-Infeld system was found by Cataldo and Garc\'{i}a in \cite{Cataldo1999} (see also Refs. \cite{Myung2008,Hendi2012}). The solution for $f\paren{r}$ is given by 
\be
f\paren{r}=-2 \beta ^2 r^2 \sqrt{1+\frac{Q^2}{\beta ^2 r^2}}+Q^2 \ln \left(1-\sqrt{1+\frac{Q^2}{\beta ^2 r^2}}\right)-Q^2 \ln \left(1+\sqrt{1+\frac{Q^2}{\beta ^2 r^2}}\right)+2 \beta ^2 r^2-\Lambda  r^2+c_1\ ,
\en
where $c_1$ is an integration constant to be determined. A reasonable criterion to determine $c_1$ is to match the asymptotic behavior with that of the Maxwell charged BTZ black hole, therefore the complete solution reads
\be\label{eq:fBIBTZ}
f\paren{r}=-m-r^2\paren{\Lambda-2\beta^2}-2r^2\beta^2\sqrt{1+\frac{Q^2}{r^2\beta^2}}+Q^2\left[1-2\ln\paren{\frac{r}{2r_0}\paren{1+\sqrt{1+\frac{Q^2}{r^2\beta^2}}}}\right]
\en
where $m$ and $Q$ are dimensionless constants that are identified as the ADM mass and charge, respectively, and it is assumed that the metric asymptotically behaves as AdS spacetime, hence $\Lambda<0$. As expected, when taking the $\beta\to\infty$ limit we obtain 
\be \label{eq:fMax}
f\paren{r}\approx-m-r^2\Lambda-2Q^2\ln\paren{\frac{r}{r_0}}+\mathcal{O}\paren{\frac{1}{\beta}}^2\ ,
\en
which is the charged BTZ solution within Maxwell electrodynamics. By setting $g_{00}=\p_rg_{00}=0$, it is possible to obtain the horizon radius for the extremal black hole as 
\be \label{eq:rexBIBTZ}
r_\ex=\frac{2Q\beta}{\sqrt{\Lambda^2-4\beta^2\Lambda}}\ .
\en

Next, in order to assess any possible singularity in the metric given by Eq. \eqref{eq:fBIBTZ}, let us analyze its behavior near the origin. As $r\to0$, one obtains
\be
f\paren{r\to0}\approx-m+Q^2\paren{1+2\ln\paren{\frac{2r_0\beta}{Q}}}-4Q\beta r+r^2\paren{2\beta^2-\Lambda}+\mathcal{O}\paren{r}^3\ .
\en
While in the Maxwell case the $g_{00}$ component diverges to $+\infty$ at the origin, for the BI-BTZ black hole the time-time component is not divergent. Notwithstanding, this solution does not represent a regular black hole since the invariants constructed from the Ricci scalar and the Ricci tensor are singular as $r\to0$, namely
\begin{align}
R&=6\Lambda-12\beta^2+4\beta\paren{\frac{r\beta^2}{\sqrt{Q^2+r^2\beta^2}}+\frac{2\sqrt{Q^2+r^2\beta^2}}{r}}\approx\frac{8Q\beta}{r}+6\Lambda-12\beta^2+\mathcal{O}\paren{r}\ , \\
R_{ab}R^{ab}&\approx\frac{24 \beta ^2 Q^2}{r^2}+\frac{32 \beta  Q \left(\Lambda -2 \beta ^2\right)}{r}+4 \left(22 \beta ^4-12 \beta ^2 \Lambda +3 \Lambda ^2\right)+\mathcal{O}\paren{r}\ .
\end{align}
\section{Thin-shell spacetime} \label{sec:TSST}
We are interested in the case of a shell of radius $R$ that divides spacetime into two parts: an inner region $\mathcal{V}^-$ for $r\le R$, and an outer region $\mathcal{V}^+$ for $r\ge R$. Since we are considering a $\paren{2+1}$-dimensional spacetime, the shell actually defines a ring. The procedure of constructing such a shell follows from applying the Darmois-Israel formalism \cite{Israel1966}. In this section we shall proceed to construct the thin-shell spacetime.

First, let us specify the inner region, $\mathcal{V}^-$, to be described by AdS spacetime, which can be regarded as the null mass BTZ spacetime, i.e.,
\be\label{eq:dsin}
\dd s^2_-=-\Lambda r^2\dd t^2+\paren{\Lambda r^2}^{-1}\dd r^2-r^2\dd\phi^2\ ,\ r\le R\ ,
\en
and the outer region, $\mathcal{V}^{+}$, corresponds to the BI-BTZ solution, i.e.,
\be
\dd s^2_+=f\paren{r}\dd t^2-f\paren{r}^{-1}\dd r^2-r^2\dd\phi^2\ , \ r\ge R\ ,
\en
with $f\paren{r}$ given by Eq. \eqref{eq:fBIBTZ}, and $x_{\pm}^a=\paren{t,r,\phi}$ are the coordinates of each region inside and outside of the shell. Both regions are glued at a timelike hypersurface, $\Sigma$, for which the line element is given in terms of the induced metric at the hypersurface, $h_{ij}$, as
\be
\dd s^2_\Sigma=h_{ij}\dd y^i \dd y^j=\dd\tau^2-R^2\paren{\tau}\dd\phi^2\ ,
\en
where $y^i=\paren{\tau,\phi}$ are the coordinates on the hypersurface and $\tau$ is the proper time of an observer located at the shell. Since the shell is located at a radius $R=R\paren{\tau}$, the $r$ and $t$ coordinates on each side of the hypersurface are parametrized by $r=R\paren{\tau}$ and $t=T\paren{\tau}$, respectively. The induced metric can be written in terms of the inner and outer spacetime metrics as
\be\label{eq:indmetric}
h_{ij}^{\pm}=g_{ab}^{\pm}e^{a}_{\pm i}e^{b}_{\pm j}\ ,
\en
where $e^{a}_{+i}$ and $e^a_{-i}$ are vectors tangent to the hypersurface viewed from the outer and inner regions, respectively.

The Darmois-Israel formalism requires the induced metric $h_{ij}$ to be continuous on the shell and the discontinuity in the extrinsic curvature to be proportional to the energy-momentum tensor of the matter on the shell. Let us denote the jump in the corresponding quantities by square brackets, thus, the first junction condition reads,
\be\label{eq:firstJC}
\left[h_{ij}\right]=0\ ,
\en
and implies that 
\be
f\paren{R}\dot{T}^2-f\paren{R}^{-1}\dot{R}^2=\Lambda R^2\dot{T}^2-\paren{\Lambda R^2}^{-1}\dot{R}^2=1\ .
\en
The second junction condition is related to the jump in the extrinsic curvature, $\left[K_{ij}\right]$ across the hypersurface, where
\be
K^{i}_{\pm j}=\paren{\nabla_a n^\pm_b} e^{a}_{\pm k}e^{b}_{\pm j}h^{ik}_{\pm}
\en
is the extrinsic curvature on each side of the hypersurface. The discontinuity in this quantity is related to the energy-momentum tensor of the shell, $S^{i}_{\ j}$, by the Lanczos equation,
\be\label{eq:Lanczos}
S^{i}_{\ j}=\frac{1}{8\pi}\paren{\left[K^{i}_{\ j}\right]-h^{i}_{\ j}\left[K\right]}\ ,
\en
where $K=h^{ij}K_{ij}$ is the trace of the extrinsic curvature.
The components for the extrinsic curvature on both sides of the shell are given by 
\begin{align}
K^{\tau}_{-\tau}&=\frac{\ddot{R}-\Lambda R}{\sqrt{\dot{R}^2+\Lambda R^2}}\ ,  \\
K^{\tau}_{+\tau}&=\frac{\ddot{R}+F'/2\dot{R}}{\sqrt{\dot{R}^2+F}}\ , \\
K^{\phi}_{-\phi}&=\frac{\sqrt{\dot{R}^2+\Lambda R^2}}{R} \ , \\
K^{\phi}_{+\phi}&=\frac{\sqrt{\dot{R}^2+F}}{F}\ ,
\end{align} 
where $F=f\paren{R}$ is defined by setting $ r=R$ in Eq. \eqref{eq:fBIBTZ} and the prime stands for the derivative w.r.t. $R$, while the dot represents differentiation w.r.t. $\tau$. Then, it is possible to compute the components of the energy-momentum tensor through Eq. \eqref{eq:Lanczos}, obtaining
\begin{align}
S^{\tau}_{\ \tau}&=\frac{\sqrt{\dot{R}^2-\Lambda  R^2}-\sqrt{-m-2 Q^2 \ln \left(\frac{R}{2r_0}(\Upsilon +1)\right)+Q^2+2 \beta ^2 (1-\Upsilon ) R^2-\Lambda  R^2+\dot{R}^2}}{8 \pi  R} \ , \label{eq:llambda} \\
S^{\phi}_{\ \phi}&=\frac{\ddot{R}-\Lambda  R}{8 \pi  \sqrt{\dot{R}^2-\Lambda  R^2}}-\frac{Q^2 \left(\ddot{R}-R \left(2 \beta ^2 \Upsilon +\Lambda \right)\right)+\beta ^2 (\Upsilon +1) R^2 \left(\ddot{R}-\Lambda  R\right)}{8 \pi  \beta ^2 (\Upsilon +1) \Upsilon  R^2 \sqrt{-m-2 Q^2 \ln \left(\frac{R}{2r_0}(1+\Upsilon) \right)+Q^2+2\beta^2 R^2\paren{1-\Upsilon}-\Lambda  R^2+\dot{R}^2}}\ , \label{eq:pp}
\end{align}
where we have defined 
\be\label{eq:Gammadef}
\Upsilon\equiv\sqrt{1+Q^2/R^2\beta^2}\ .
\en
 On the other hand, we can consider that the matter in the shell can be modeled as a perfect fluid with an energy-momentum tensor given by
\be
\mathcal{S}^{i}_{\ j}=\paren{\lambda+p}u^{i}u_j-ph^{i}_{\ j}\ ,
\en
where $\lambda$ is the linear energy density of the ring and $p$ is the tangential pressure. Consequently, we have $S^{\tau}_{\ \tau}=\lambda$ and $S^{\phi}_{\ \phi}=-p$. Imposing that the shell is static, i.e., $R=R_0$ and $\dot{R}=\ddot{R}=0$ leads to the equations for the energy density and pressure for the static ring
\begin{align}
\lambda_0&=\frac{\sqrt{-\Lambda R_0^2}-\sqrt{-m-2 Q^2 \ln \left(\frac{R_0}{2r_0}(1+\Upsilon_0) \right)+Q^2+2 \beta ^2 (1-\Upsilon_0) R_0^2-\Lambda  R_0^2}}{8 \pi  R_0} \ , \label{eq:lambda0}\\
p_0&= -\frac{1}{8\pi}\left[\sqrt{-\Lambda }+\frac{Q^2 \left(2 \beta ^2 \Upsilon_0+\Lambda \right)+\beta ^2 (1+\Upsilon_0) \Lambda  R_0^2}{\beta ^2 (\Upsilon_0+1) \Upsilon_0 R_0 \sqrt{-m-2 Q^2 \ln \left(\frac{R_0}{2r_0}(1+\Upsilon_0)\right)+Q^2+2\beta^2R_0^2\paren{1-\Upsilon_0}-\Lambda  R_0^2}}\right]\ .\label{eq:p00}
\end{align}
In the latter, all quantities with a subindex $0$ indicate that they are evaluated at the equilibrium configuration, $R_0$. The proper mass of the shell is related to the linear energy density by the relation
\be \label{eq:lambda}
M=2\pi R_0 \lambda_0\ .
\en
In order to simplify our expressions, let us write the cosmological constant as $\Lambda=-l^{-2}$, where $l$ is the cosmological length. Likewise, let us note that the constant $r_0$ appearing in the BI-BTZ metric (cf. Eq. \eqref{eq:fBIBTZ}) ensures that the correct asymptotic behavior for the electrostatic potential is taken into account when computing this component. Thus, as $l$ defines  a characteristic length scale of the spacetime, we can set $r_0=l$. In virtue of Eq. \eqref{eq:lambda}, we can express the ADM mass in terms of $\paren{M,Q,\beta,R_0}$ as
\be \label{eq:mMRQ}
m=8\frac{MR_0}{l}-16M^2+2R_0^2\beta^2\paren{1-\Upsilon_0}+Q^2\paren{1-2\ln\left[\frac{R_0}{2l}\paren{1+\Upsilon_0}\right]}\ ,
\en
which allows us to write the energy density and  pressure as 
\be \label{eq:lambdacorr}
\lambda_0=\frac{R_0-\sqrt{\paren{R_0-4lM}^2}}{8\pi l R_0}=\frac{M}{2\pi R_0}\ , 
\en
\be \label{eq:p0}
p_0=\frac{2 (\Upsilon_0+1) M R_0-l Q^2}{4 \pi  (\Upsilon_0+1) R_0 (R_0-4 l M)}\ . 
\en
Notice that in Eq. \eqref{eq:lambdacorr} we have used the fact that $4lM< R_0$, which can be regarded as a primary bound on $M$ or on the compactness of the shell.

It is possible to express the shell's charge in terms of a linear charge density following the approach of Kucha{\v r} (see \cite{Kuchar1968,Lemos2015}) which consist in obtaining the junction conditions for the electromagnetic field and imply that the tangential components of the field across the shell must be zero and the normal components are related to the surface current and are expressed as $\left[F_{ij}\right]=0$ and $\left[E_{i\perp}\right]=2\pi\lambda_e u_i$, respectively, where $\lambda_e$ is the linear charge density and $u_i$ is the observer's velocity. Notice that in NLED, the source generates the excitation field $E_{ij}$ which, in turn, gives the components of the electric field. Therefore, the total charge of the shell can be expressed in terms of the linear charge density as
\be\label{eq:lambdae} 
Q=2\pi R_0\lambda_e\ .
\en

\section{Dynamical stability} \label{sec:dynstab}
The linear energy density, $\lambda$, and the pressure, $p$, obtained from Eqs. \eqref{eq:llambda} and \eqref{eq:pp} satisfy the conservation equation
\be\label{eq:cons}
\frac{d}{d\tau}\paren{R\lambda}+p\frac{dR}{d\tau}=0 \ ,
\en
where $R$ is the radius of the ring. This equation can also be written as 
\be \label{eq:dlambda}
\dot{\lambda}=-\paren{\lambda+p}\frac{\dot{R}}{R}\ ,
\en
which will be useful when exploring the shell dynamics. From Eq. \eqref{eq:llambda} it follows that the dynamics of the ring is given by the equation
\be \label{eq:dReqn}
\dot{R}^2+V\paren{R}=0\ ,
\en
where the potential $V\paren{R}$ is given by 
\begin{align}
V\paren{R}&=-\frac{1}{2}\left[m-Q^2\paren{1-2\ln\left[\frac{R}{2l}\paren{1+\Upsilon}\right]}\right]-\paren{4\pi R \lambda}^2-\frac{1}{\paren{16\pi R\lambda}^2}\left[m-Q^2\paren{1-2 \ln \left[\frac{R}{2l}(\Upsilon +1)\right]}\right]^2 \nonumber\\
&-\frac{\beta^2\paren{\Upsilon-1}}{\paren{8\pi \lambda}^2}\paren{m+2Q^2\ln\left[\frac{R}{2l}\paren{\Upsilon+1}\right]-\beta^2R^2\Upsilon\paren{\Upsilon-1}}-\left[\beta^2\paren{\Upsilon-1}+\Lambda\right]R^2\ . \label{eq:V}
\end{align}
Equation \eqref{eq:dReqn} allows for the study of radial perturbations around the equilibrium configuration. Assuming that the equilibrium state is attained when $R=R_0$, with energy density and pressure, $\lambda_0$ and $p_0$ given by Eqs. \eqref{eq:lambda0} and \eqref{eq:p00}, respectively. In order to study the dynamical stability, let us expand the potential \eqref{eq:V} around the static configuration as
\be
V\paren{R}=V\paren{R_0}+V'\paren{R_0}\paren{R-R_0}+\frac{V''\paren{R_0}}{2}\paren{R-R_0}^2+\mathcal{O}\paren{R-R_0}^3\ .
\en

The first and second derivatives of the potential can be written as
\begin{align}
V'(R)&=\frac{\lambda'  \left(4 \beta ^2 (\Upsilon -1) R^2 \left(2 Q^2 g+m-\beta ^2 (\Upsilon -1) \Upsilon  R^2\right)+\left(m-Q^2 \left(1-2g \right)\right)^2-(8 \pi  R \lambda)^4\right)}{128 \pi ^2 R^2 \lambda^3}\nonumber\\
&+\frac{\left(m-Q^2 \left(1-2g \right)\right)^2-4 \beta ^4 (\Upsilon -1)^2 R^4-(8 \pi  R \lambda)^4}{128 \pi ^2 R^3 \lambda^2}-2 R \left(\beta ^2 (\Upsilon -1)+\Lambda \right)\ ,
\end{align}
\begin{align} \label{eq:VppBI}
V''\paren{R}&=-Q^2g''-\beta^2\paren{2\paren{\Upsilon-1}+4R\Upsilon'+R^2\Upsilon''}-2\Lambda-\frac{4}{\paren{16\pi}^2\paren{R\lambda}^2}\paren{2\paren{Q^2g'}^2+Q^2g''}\\
&+\frac{6}{\paren{16\pi}^2}p\paren{\frac{2}{\paren{R\lambda}^3}Q^2g'-\frac{p}{\paren{R\lambda}^4}}\paren{m-Q^2+2Q^2g}\nonumber\\
&+\frac{\beta^2}{\paren{8\pi}^2\lambda^2}\paren{-\Upsilon''+\left[\frac{4}{R^2}\paren{\Upsilon-1}-\frac{1}{R}\paren{4\Upsilon'-\frac{6}{\lambda}\paren{\Upsilon-1}}\right]\paren{1+\frac{p}{\lambda}}}\paren{m+2Q^2g-\beta^2R^2\Upsilon\paren{\Upsilon-1}}\nonumber\\
&-\frac{\beta^2}{\paren{8\pi}^2\lambda^2}\paren{\frac{4}{R}\paren{\Upsilon-1}\paren{1+\frac{p}{\lambda}}+\Upsilon'}\paren{2Q^2g'-2\beta^2R\Upsilon\paren{\Upsilon-1}+\beta^2R^2\Upsilon'\paren{1-2\Upsilon}}\nonumber\\
&-\frac{\beta^2}{\paren{8\pi}^2\lambda^2}\paren{\Upsilon-1}\paren{2Q^2g''-2\beta^2\Upsilon\paren{\Upsilon-1}+\beta^2\paren{1-2\Upsilon}\paren{4R\Upsilon'+R^2\Upsilon''}-2\beta^2R^2\paren{\Upsilon'}^2} \nonumber\\
&+\frac{2}{R^2}\paren{1+\frac{p}{\lambda}}\left[\frac{1}{\paren{16\pi}^2\paren{R\lambda^2}}\paren{m-Q^2+2Qg}^2-16\pi^2R\lambda+\frac{\beta^2}{\paren{8\pi}^2\lambda^2}\paren{\Upsilon-1}\paren{m+2Q^2g-\beta^2R^2\Upsilon\paren{\Upsilon-1}}\right]\frac{p'}{\lambda'} \nonumber
\end{align}
where
\begin{align}
\Upsilon&=\sqrt{1+\frac{Q^2}{\beta^2R^2}}\ , \\
\Upsilon'&=-\frac{1}{\Upsilon R}\paren{\Upsilon-1}\paren{\Upsilon+1}\ , \\
\Upsilon''&=\frac{\paren{\Upsilon+1}^2\paren{\Upsilon-1}^2}{\Upsilon^3 R^2}\ , \\
g&=\ln\paren{\frac{R}{2l}\paren{\Upsilon+1}}\ , \\
g'&=\frac{1+\Upsilon+R\Upsilon'}{R\paren{\Upsilon+1}}\ ,\\
g''&=\frac{-\paren{\Upsilon+1+R\Upsilon'}^2+R\paren{\Upsilon+1}\paren{2\Upsilon'+R\Upsilon''}}{R^2\paren{\Upsilon+1}^2} \ .
\end{align}
In virtue of Eq. \eqref{eq:dlambda}, we can write $\lambda'=-\paren{\lambda+p}/R$ and $\lambda''=-\paren{\lambda'+p'}/R+\paren{\lambda+p}/R^2$. Straightforward computation shows that $V\paren{R_0}=V'\paren{R_0}=0$, and the stability condition can be resumed as $V''\paren{R_0}>0$.

While the stability criterion in the latter paragraph is derived from the equation of motion for the shell (see Eq. \eqref{eq:dReqn}), imposing the conditions $V\paren{R_0}=0$ and the existence of a minimum of the potential for this configuration imply that the squared shell velocity becomes negative in the vicinity of the equilibrium configuration. This seemingly unphysical behavior can be avoided by considering a shift in the potential of the form $V\paren{R}\to V\paren{R}-\epsilon^2$, with $\epsilon\in\mathbb{R}
$. This guarantees that the minimum of the potential corresponds to zero velocity and also that $\dot{R}\in\mathbb{R}$ in the vicinity of this minimum, which corresponds to a \emph{bounded excursion} as introduced in Ref. \cite{Visser2004}.\footnote{We thank the anonymous referee for pointing out this important detail regarding dynamical stability.}

\subsection{The $\beta\to\infty$ limit}
\begin{figure}[ht]
\includegraphics[width=0.47\linewidth]{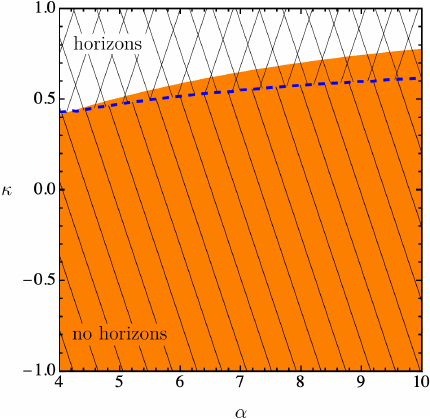}
\caption{(Orange region) Dynamical stability for a RN ring obeying a linear equation of state. The upper meshed region represents the region in the parameter space where the solution allows for two horizons for the outer manifold. The lower meshed region represents configurations in which the shell is glued over an overcharged spacetime (i.e., no horizons exist). The blue dashed line represents the extremal configuration where there is only one horizon.}
\label{fig:RNBTZdynstab}
\end{figure}

Let us first explore the simpler case of the shell constituted of matter obeying Maxwell electrodynamics. This corresponds to the BI-BTZ case where $\beta\to\infty$. In this case, the potential \eqref{eq:V} reads
\be
V\paren{R}=-\frac{1}{2}\paren{m+2Q^2\ln\left[\frac{R}{l}\right]-2\frac{R^2}{l^2}}-\paren{4\pi R\lambda}^2-\frac{1}{\paren{16\pi R\lambda}^2}\paren{m+2Q^2\ln\left[\frac{R}{l}\right]}^2\ .
\en
The potential and its first derivative evaluated at $R_0$ take the simple form
\begin{align} \label{eq:potRNR0}
V\paren{R_0}&=-\frac{\left(M^2-4 \pi ^2 \lambda_0^2 R_0^2\right) \left(4 l^2 \left(M^2-4 \pi ^2 \lambda_0^2 R_0^2\right)-4 l M R_0+R_0^2\right)}{4 \pi ^2 \lambda_0^2 l^2 R_0^2}\ , \\
V'\paren{R_0}&=-\frac{M^2 p_0}{2 \pi ^2 \lambda_0^3 l^2 R_0}+\frac{2 R_0}{l^2}+\frac{M \left(16 M^2 p_0-\lambda_0 Q^2\right)}{8 \pi ^2 \lambda_0^3 l R_0^2}-\frac{\left(M^2-4 \pi ^2 \lambda_0^2 R_0^2\right) \left(8 M^2 p_0+\lambda_0 \left(32 \pi ^2 \lambda_0 p_0 R_0^2-Q^2\right)\right)}{4 \pi ^2 \lambda_0^3 R_0^3}\ .
\end{align}
Considering that the energy density and pressure for the equilibrium state are given by
\begin{align}
\lambda_0&=\frac{M}{2\pi R_0}\ , \label{eq:lambdaM}\\
p_0&=
\frac{-l^2 Q^2-R_0 \sqrt{(R_0-4 l M)^2}+R_0^2}{8 \pi  l R_0 \sqrt{(R_0-4 l M)^2}}\ ,
\end{align}
we obtain directly that $V\paren{R_0}=0$ and $V'\paren{R_0}=0$. Next, noticing that the shell's mass and charge are given by $M=2\pi R_0\lambda$ and $Q=2\pi R_0\lambda_e$, it is possible to relate both the material and charge densities as $\lambda_e=\alpha \lambda_0$. Hence, the parameter $\alpha$ corresponds to the charge-to-mass ratio for the shell. As shall be noted below, this parametrization is useful for studying the stability configurations. Next, we can express the relevant quantities as functions of $\paren{\lambda_0,\alpha}$. Namely, for the pressure we have,
\be\label{eq:pressure0}
p_0=
-\frac{4 \pi ^2 \alpha ^2 \lambda_0 ^2 l^2+\sqrt{(8 \pi  \lambda_0  l-1)^2}-1}{8 \pi  l \sqrt{(1-8 \pi  \lambda_0  l)^2}}\ .
\en
It is interesting to note that, within this parametrization, $p_0$ is independent of $R_0$. Furthermore, we can define the dimensionless variable $y\equiv\lambda_0l$, such that the pressure in Eq. \eqref{eq:pressure0} can be written as
\be\label{eq:p0y}
p_0=-\frac{\lambda_0 \left(4 \pi ^2 \alpha ^2 y^2+\sqrt{(8 \pi  y-1)^2}-1\right)}{8 \pi  y \sqrt{(1-8 \pi  y)^2}}\ .
\en
Noticing that from Eq. \eqref{eq:lambdacorr}, we have that $y<1/8\pi$, the latter equation reduces to 
\be \label{eq:p0y1}
p_0=\frac{\lambda_0 \left(\pi  \alpha ^2 y-2\right)}{16 \pi  y-2}\ .
\en
 Additionally, the second derivative of the potential is now written as
\be \label{eq:VppRN}
V''\paren{R_0}=-\frac{\lambda_0\left(\lambda_0 \left(4 \pi ^3 \alpha ^2 \left(\alpha ^2-16\right) y^3+16 \pi ^2 \alpha ^2 y^2-\pi  \left(3 \alpha ^2+16\right) y+4\right)-2 R_0 (8 \pi  y-1)^3 p'(R_0)\right)}{y^2 (1-8 \pi  y)^2}\ .
\en
In order to evaluate the dynamical stability we need to specify an equation of state. We shall parametrize a barotropic equation of state of the form
\be
p=x\paren{\lambda}\lambda\ ,
\en
in order to evaluate the stability regions. For a generic equation of state, from Eq. \eqref{eq:p0y1} we obtain $y$ to be
\be
y=\frac{2 \paren{p_0- \lambda_0}}{16 \pi  p_0-\pi  \alpha ^2 \lambda_0}<\frac{1}{8\pi} .
\en
Thus,  for a linear equation of state we can set $x\paren{\lambda}=\kappa$ and we have
\be
0\le\frac{2 \paren{\kappa- 1}}{\pi\paren{16   \kappa-  \alpha ^2 }}<\frac{1}{8\pi} .
\en
The latter inequalities can be solved for $\paren{\alpha,\kappa}$. Thus, the corresponding regions in the $\paren{\alpha,\kappa}$ plane where $y$ is meaningful are
\be \label{eq:ymeaning}
\begin{cases}
\alpha>4\qquad ,\ \text{for}\ \kappa<1 \\ 
0<\alpha\qquad , \ \text{for}\ \kappa=1 \\
0<\alpha<4\ , \ \text{for}\ \kappa>1\ .
\end{cases}
\en
Next, plugging the solution for $y$ into Eq. \eqref{eq:VppRN} allows us to determine the dynamical stability regions for such a solution. These conditions must be matched with the regions where $y$ is meaningful (given by Eq. \eqref{eq:ymeaning}) and contrasted with those regions where the solution does not describe a physically viable situation, such as the appearance of imaginary solutions for the horizons and regions where the Weak Energy Condition (WEC) is not fulfilled. These conditions restrict the allowable parameter space to $\alpha>4$. Figure \ref{fig:RNBTZdynstab} displays the dynamical stability regions for a RN ring obeying a linear equation of state. It can be seen that both regions in the parameter space (with and without horizons enclosed by the shell) possess dynamical stability regions. Moreover, considering that the matter obeys a linear equation of state, we have been able to reduce the stability analysis to configurations in the $\paren{\alpha,\kappa}$ plane. 
\subsection{The BI-BTZ shell}

In the case where $\beta$ is finite, the condition for dynamical stability, $V''\paren{R_0}>0$, needs to be evaluated using the function in Eq. \eqref{eq:VppBI}. Hence, for a linear equation of state we obtain
\begin{align} \label{eq:VppcondBI}
V''\paren{R_0}&=\frac{Q^2}{\beta ^2 \Upsilon_0 (\Upsilon_0+1)^2 l^2 M R_0^2 (R_0-4 l M)^2} \left(-\frac{2 l^3 Q^2 \left(\Upsilon_0 Q^2-8 (\Upsilon_0+1) M^2\right)}{(\Upsilon_0-1) (\Upsilon_0+1) R_0}-\frac{8 l^2 M Q^2}{\Upsilon_0-1}-\right. \nonumber\\
&\left.-\frac{\Upsilon_0 \kappa  Q^2 (R_0-4 l M)^2 \left(l \left(8 (\Upsilon_0+1) M^2+Q^2\right)-4 (\Upsilon_0+1) M R_0\right)}{\beta ^2 (\Upsilon_0-1)^2 (\Upsilon_0+1) R_0^3}+\frac{l R_0 \left(8 \Upsilon_0 (\Upsilon_0+1) M^2+(2 \Upsilon_0+1) Q^2\right)}{\Upsilon_0-1}-\right. \nonumber\\
&\left.-\frac{8 M R_0^2}{\Upsilon_0^2-1}-\frac{8 \Upsilon_0 M R_0^2}{\Upsilon_0^2-1}-4 \Upsilon_0 M R_0^2-8 M R_0^2\right)>0\ .
\end{align}
In a similar fashion as in the Maxwell case, the electromagnetic junction condition, Eq. \eqref{eq:lambdae}, guarantees that the shell's charge can be written in terms of a linear charge density, while the material mass is related to the linear energy density as $M=4\pi R_0\lambda_0$. Thus, the parametrization for the charge-to-mass relation used before, $Q=\alpha M$ can still be used leading to $\lambda_e=\alpha\lambda_0$, and, in principle, it should be still possible to study the stability condition in the $\paren{\alpha,\kappa}$ plane. Yet, it is expected that the BI constant, $\beta$ will explicitly appear in some combination of the parameters and, just as in the latter case, we shall use the pressure equation of state in order to constrain the allowable configurations.  

\begin{figure}[ht]
\includegraphics[width=0.87\linewidth]{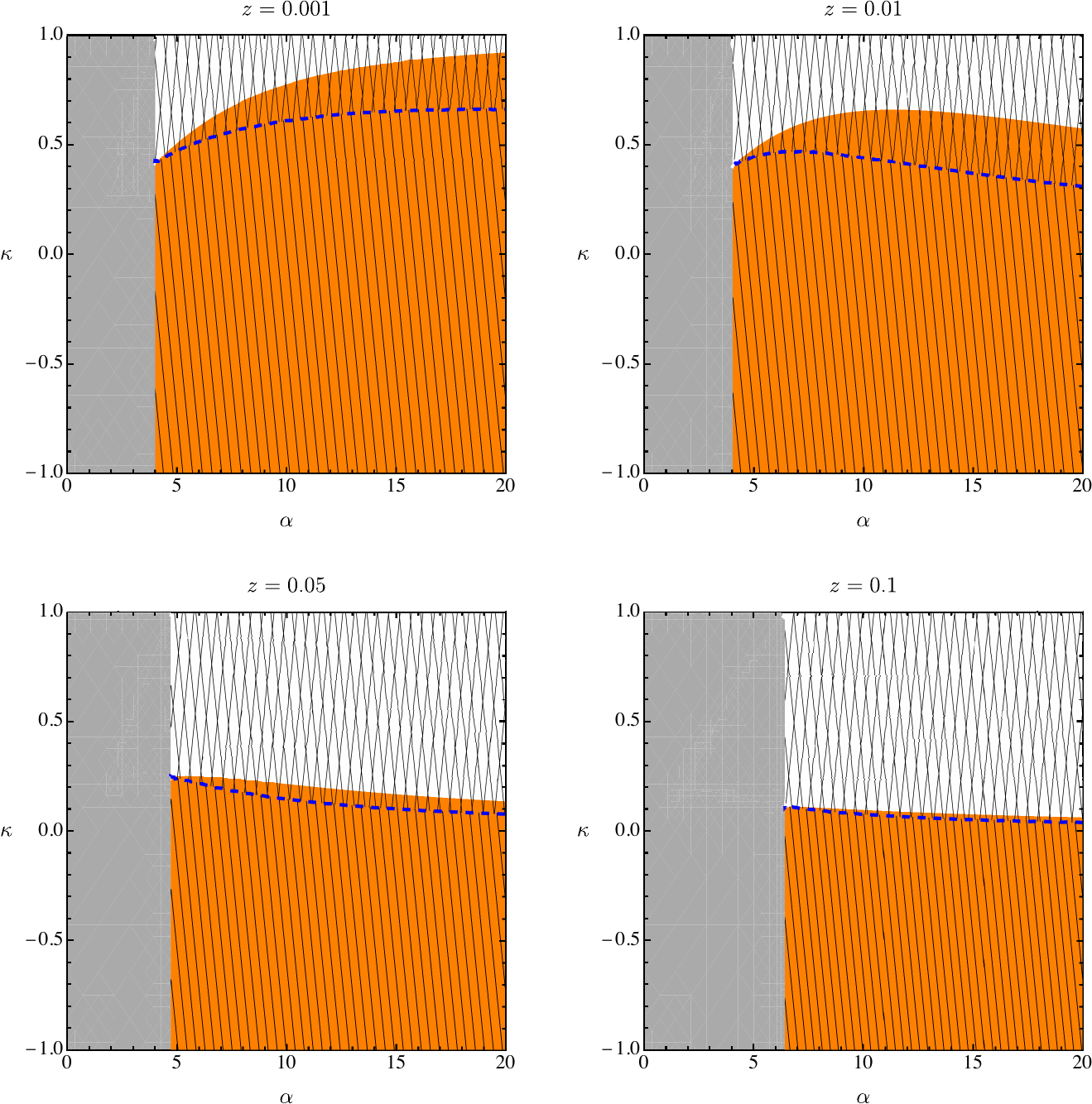}
\caption{Regions of dynamical stability for a $\paren{2+1}$-dimensional BI thin-shell for varying $z=\lambda/\beta$. Notice that as $z$ grows the stability regions are shifted towards the domain where the outer spacetime is overcharged.}
\label{fig:BIBTZdynstab}
\end{figure}

Turning to the pressure for the equilibrium configuration in Eq.~\eqref{eq:p00} and defining $y\equiv\lambda_0l$ and $z\equiv\lambda_0/\beta$, we obtain that for a linear equation of state
\be
\kappa=-\frac{\left(-\pi  \alpha ^2 y+\sqrt{4 \pi ^2 \alpha ^2 z^2+1}+1\right)}{(8 \pi  y-1) \left(\sqrt{4 \pi ^2 \alpha ^2 z^2+1}+1\right)}\ ,
\en
which can be solved for $y$ leading
\be \label{eq:ysolBI}
y=\frac{\paren{\kappa-1}\paren{1+\sqrt{4\pi^2\alpha^2z^2+1}}}{-\pi  \alpha ^2+8 \pi  \kappa \paren{1+  \sqrt{4 \pi ^2 \alpha ^2 z^2+1}}}\ .
\en
Considering the primary bound $0\le y<1/8\pi$ we obtain that the meaningful region where the matter respects the WEC is given by 
\be \label{eq:condalphaBI}
\alpha>\sqrt{16+256 \pi^2 z^2}\ .
\en
Plugging Eq. \eqref{eq:ysolBI} into \eqref{eq:VppcondBI} leads to a condition solely dependent on the parameters $\paren{\alpha,\kappa,z}$ and can be evaluated for fixed $z$. Fig. \ref{fig:BIBTZdynstab} displays the dynamical stability regions for the BI shell for several values of $z$. In the Figure we also display the curve that delimits the parameters of the thin-shell solution such that the outer manifold represents a BI-BTZ solution that resembles a black hole or a naked singularity, i.e., the blue curves represent the situation where the outer solution corresponds to an extremal BI-BTZ configuration. The parameter $z$ controls the departures from Maxwell electrodynamics and $z\to0$ recovers the dynamical stability for the Maxwell case. As $z$ grows the bound \eqref{eq:condalphaBI} increases the minimum possible value that $\alpha$ can take and the horizon structure drastically changes. Notice that already for $z=0.1$, the stability configurations correspond to the extremal and overcharged domain for the thin shell solutions, which shall prevent us from considering a Hawking type entropy for studying the thermodynamical stability later on.
\section{Thermodynamics of the BI-BTZ shell} \label{sec:THshell}
We begin with the thermodynamic description of a thin shell embedded in a $\paren{2+1}-$ dimensional spacetime whose entropy is characterized by its mass $M$, its charge $Q$ and its perimeter $P$. Thus, we can write the first law of thermodynamics for the shell as
\be 
TdS=dM+p dP-\Phi dQ\ .
\en
Alternatively, defining $\beta_T\equiv1/T$, we can write 
\be \label{eq:firstlaw}
dS=\beta_T dM+\beta_T p dP-\beta_T\Phi dQ\ .
\en
In order for the entropy differential to be exact, we have the following integrability conditions.
\begin{align}
\paren{\frac{\p\beta_T}{\p R}}_{M,Q}&=2\pi\paren{\frac{\p\beta_T p}{\p M}}_{R,Q}\ , \label{eq:intcond1}\\
\paren{\frac{\p\beta_T}{\p Q}}_{M,R}&=-\paren{\frac{\p\beta_T\Phi}{\p M}}_{R,Q}\ ,\label{eq:intcond2} \\
\paren{\frac{\p\beta_T\Phi}{\p R}}_{M,Q}&=-2\pi\paren{\frac{\p\beta_T p}{\p Q}}_{M,R}\ . \label{eq:intcond3} 
\end{align}

We can define the redshift factor for the shell as $k\equiv\sqrt{f\paren{R}}$ where $f(R)$ is the metric function for the spacetime in the exterior of the shell. In particular, we have that the material mass can be written in terms of $k$ as
\be\label{eq:Mk}
M=\paren{\frac{R}{4l}-\frac{k}{4}}\ .
\en
Since $k\ge0$, the latter equation implies directly that $4lM\le R$, which is the bound previously obtained. From the above expression it is straightforward to obtain
\be \label{eq:dMdR}
\paren{\frac{\p M}{\p R}}_{m,Q}=-2\pi p\ .
\en
These two latter expressions are general and independent of the functional form of the exterior spacetime and shall be useful when treating the thermodynamics of charged shells. Once the entropy in Eq. \eqref{eq:firstlaw} depends on the intensive parameters $\paren{\beta_T,p,\Phi}$, we need to specify an equation of state for each of these quantities. The pressure equation of state is readily given by the junction conditions and takes the form displayed in Eq. \eqref{eq:p00}. Therefore, the remaining task is to specify an equation of state for the inverse temperature, $\beta_T$ and the thermodynamic electrostatic potential, $\Phi$.

It is possible to express the quantities $M$ and $Q$ as implicit functions of the horizons $r_+$ and $r_-$, such that the temperature equation of state can be obtained by solving Eq. \eqref{eq:intcond1}, which reads
\be \label{eq:tos}
\paren{\frac{\p\beta_T}{\p R}}_{r_+,r_-}=2\pi\beta_T\paren{\frac{\p p}{\p M}}_{R,Q}\ .
\en
Next, the electrostatic potential equation of state can be obtained by using Eq. \eqref{eq:dMdR} into the integrability conditions \eqref{eq:intcond1}-\eqref{eq:intcond3}, which leads to the equation
\be
k\paren{\frac{\p p}{\p Q}}_{M,R}+\frac{1}{2\pi}\paren{\frac{\p \Phi k}{\p R}}_{r_+,r_-}=0\ .
\en
Once all the equations of state are known, it is possible to obtain the entropy of the shell. Next, we analyze the thermodynamics for a charged thin-shell in a $\paren{2+1}$-dimensional spacetime both in Maxwell and Born-Infeld electrodynamics.

\subsection{Maxwell charged thin shell}
Assuming that the exterior part of the shell is described by the $\paren{2+1}$-dimensional solution of the Einstein-Maxwell system (i.e., by Eq. \eqref{eq:fMax}), we have that the pressure of the shell is given by
\be
p=\frac{R\paren{R-kl}-l^2Q^2}{8\pi k l^2 R}\ .
\en
Next, the temperature equation of state can be obtained by solving Eq. \eqref{eq:tos}, which reads
\be
\paren{\frac{\p\beta_T}{\p R}}_{r_+,r_-}=\beta_T\paren{\frac{R^2-l^2Q^2}{k^2l^2R}}\ ,
\en
and has the solution $\beta_T={\rm a}\paren{r_+,r_-}k$, where ${\rm a}\paren{r_+,r_-}$ can be envisaged as the inverse temperature of the shell if its radius were $R=l$. Next, for the thermodynamic electrostatic potential we obtain the equation
\be
\paren{\frac{\p \Phi k}{\p R}}_{M,R}=\frac{Q}{2R}\ ,
\en
which has the solution
\be \label{eq:PhiMax}
\Phi=\frac{1}{k}\paren{{\bar\phi}\paren{r_+,r_-}+\frac{Q}{2}\ln\paren{\frac{R}{l}}}\ .
\en
Again, the function ${\bar \phi}\paren{r_+,r_-}$ is regarded as the electrostatic potential of the shell if its radius were $l$ and shall be treated as an unknown function in order to determine the integrability conditions. Therefore, the thermodynamic electrostatic potential, $\Phi$ can be thought of as the blueshifted difference between the electrostatic potential of a shell of radius $R$ and that of a shell of radius $R=l$. In the following, it is useful to write the thermodynamic electrostatic potential as 
\be\label{eq:PhiMaxC}
\Phi=\frac{Q}{2k}\left[{\rm c}\paren{r_+,r_-}+\ln\paren{R/l}\right]\ ,
\en   
where ${\rm c}\paren{r_+,r_-}\equiv2{\bar \phi}\paren{r_+,r_-}/Q$.

Turning to the entropy differential in Eq. \eqref{eq:firstlaw}, it is always possible to write it in terms of two implicit functions $r_+\paren{M,Q}$ and $r_-\paren{M,Q}$ which are related to the roots of the metric function. Notice that $M=M\paren{m,Q,R}$ is given by Eq. \eqref{eq:Mk}, while $Q$ is the charge of the shell, that is not related to its radius. Hence, straightforward calculation shows that the entropy differential can now be written in terms of $\dd r_+$ and $\dd r_-$ alone as
\be \label{eq:dSrprm}
\dd S=\beta_T\left[\paren{\frac{\p M}{\p r_+}}_{r_-,R}-\Phi\paren{\frac{\p Q}{\p r_+}}_{r_-,R}\right]\dd r_++\beta_T\left[\paren{\frac{\p M}{\p r_-}}_{r_+,R}-\Phi\paren{\frac{\p Q}{\p r_-}}_{r_+,R}\right]\dd r_-\ .
\en
The remaining task being to compute the derivatives inside the square brackets. Notice that the metric function in Eq. \eqref{eq:fMax} is non-polynomial and the logarithmic term prevents us from obtaining a linear function of the horizons for $m$ and $Q$ as in the case of the Reissner-Nordstr\"{o}m black hole in 3+1 dimensions. However, it is still possible to analytically find a relation for $m$ and $Q$ as functions of $\paren{r_+,r_-}$. Taking $f\paren{r_+}=0$ and $f\paren{r_-}=0$ we find that the charge, the ADM mass and the material mass are
\begin{align}
Q^2&=\frac{r_-^2-r_+^2}{2l^2\ln\paren{r_-/r_+}}\ , \\
m&=
\frac{1}{2} \left(\frac{r_-^2+r_+^2}{l^2}-\frac{\left(r_-^2-r_+^2\right) \ln \left(\frac{r_- r_+}{l^2}\right)}{l^2 \ln \left(\frac{r_-}{r_+}\right)}\right)\ ,\\ 
M&=
\frac{R}{4 l}-\frac{1}{4} \sqrt{\frac{R^2}{l^2}-\frac{\left(r_-^2-r_+^2\right) \ln \left(\frac{R}{l}\right)}{l^2 \ln \left(\frac{r_-}{r_+}\right)}+\frac{1}{2} \left(\frac{\left(r_-^2-r_+^2\right) \ln \left(\frac{r_- r_+}{l^2}\right)}{l^2 \ln \left(\frac{r_-}{r_+}\right)}-\frac{r_-^2+r_+^2}{l^2}\right)}\ , 
\end{align}
respectively. Notice that as $r_-\to0$ we recover the results for the BTZ black hole. Thus, computing the corresponding derivatives in Eq. \eqref{eq:dSrprm} results in a new integrability condition for guaranteeing that the entropy differential is exact, namely

\be \label{eq:intcondM}
  \frac{\p {\rm a}}{\p r_-}h(r_+)r_- \left[{\rm c}+\ln \left(\frac{r_-}{l}\right)\right]+\frac{\p {\rm a}}{\p r_+}h(r_-)r_+   \left[{\rm c}+\ln \left(\frac{r_+}{l}\right)\right]=-{\rm a} \left(\frac{\p {\rm c}}{\p r_-} h(r_+) r_-+\frac{\p {\rm c}}{\p r_+} h(r_-) r_+\right)\ ,
\en
where ${\rm a}={\rm a}\paren{r_+,r_-}$ and ${\rm c}={\rm c}\paren{r_+,r_-}$, and we have defined
\be
h\paren{r_{\pm}}\equiv r_+^2-r_-^2+2r_\pm^2\ln\paren{r_-/r_+}\ .
\en

One may attempt to find a family of solutions for ${\rm a}\paren{r_+,r_-}$ and ${\rm c}\paren{r_+,r_-}$ that satisfy the integrability condition. However, we shall propose an {\em ansatz} for the inverse temperature that resembles the Hawking temperature for a charged BTZ black hole \cite{Medved2002}, namely
\be \label{eq:aHawking}
{\rm a}\paren{r_+,r_-}=\frac{\gamma  l^2 r_+ \ln \left(\frac{r_-}{r_+}\right)}{r_+^2-r_-^2+2 r_+^2 \ln \left(\frac{r_-}{r_+}\right)}\ ,
\en
where $\gamma$ is a constant to be determined by the matter content of the shell (i.e., the microphysics). Plugging this expression into the integrability condition in Eq. \eqref{eq:intcondM} still results in an equation for ${\rm c}\paren{r_+,r_-}$ and its derivatives. We may propose an {\em ansatz} for this function of the form
\be \label{eq:ansatzCM}
{\rm c}\paren{r_+,r_-}=-\ln\paren{r_+/l}\ ,
\en
that fulfills the integrability condition. Notice that the sign in Eq. \eqref{eq:ansatzCM} is such that the thermodynamic electrostatic potential in Eq. \eqref{eq:PhiMax} is effectively a difference of potentials. With these \emph{ans\"{a}tze} at hand, the entropy differential is 
\be
\dd S=\frac{\gamma}{8}\dd r_+\ ,
\en
which can be directly integrated yielding $S=S_0+\gamma r_+/8$, with $S_0$ being an integration constant. This constant can be fixed by demanding that in the absence of a shell (i.e., when $M$ and $Q$ go to zero), the entropy must also be zero, hence $S_0=0$ and the entropy for the shell is 
\be \label{eq:entropyMax}
S=\frac{\gamma}{8}r_+\ .
\en
where $\gamma>0$ necessarily in order to have a positive entropy. The constant $\gamma$ can take, \emph{a priori}, arbitrary values. However, in the particular limit where the shell's radius approaches the gravitational radius, i.e., as $R\to r_+$, one must recover the Hawking temperature for the shell and hence the entropy will be that of the corresponding black hole \cite{Lemos2015}. Hence, in this limit we have 
\be
\gamma=4\pi\ .
\en

\subsection{Born-Infeld charged thin shell}

Let us now focus on the more general case of a BI charged thin shell. In this case, the pressure of the thin shell is given by Eq. \eqref{eq:p00} and the remaining equations of state come from solving the integrability conditions. Namely, the first integrability condition can be written as
\begin{align}
\paren{\frac{\p\beta_T}{\p R}}_{r_+,r_-}&=2\pi\beta_T\paren{\frac{\p p}{\p M}}_{R,Q} \\ 
&=\beta_T\paren{\frac{R^2\beta^2\paren{1+\Upsilon}+Q^2\paren{1-2l^2\Upsilon\beta^2}}{Rk^2l^2\beta^2\Upsilon\paren{1+\Upsilon}}}\ .
\end{align}
The latter equation has a general solution of the form $\beta_T=\mathrm{a}\paren{r_+,r_-}k\paren{R,r_+,r_-}$, where $\mathrm{a}=\mathrm{a}\paren{R\to l,r_+,r_-}$ can be envisaged as the inverse temperature of the shell if its radius were $l$.

For the electrostatic potential, the integrability conditions in Eqs. \eqref{eq:intcond1}-\eqref{eq:intcond3} can be expressed as
\be
\frac{1}{2\pi}\paren{\frac{\p\Phi}{\p R}}_{r_+,r_-}+\Phi\paren{\frac{\p p}{\p M}}_{R,Q}+\paren{\frac{\p p}{\p Q}}_{M,R}=0\ .
\en
In use of Eq. \eqref{eq:dMdR}, and expressing the corresponding derivatives, it is possible to write the latter equation as
\be
2R\Upsilon\paren{\frac{\p\paren{\Phi k}}{\p R}}_{r_+,r_-}-Q=0\ ,
\en
which integrates as
\be \label{eq:PhiBI}
\Phi\paren{R,r_+,r_-}=\frac{1}{k}\paren{{\bar \phi}\paren{r_+,r_-}-\frac{1}{4}Q\ln\paren{\frac{\Upsilon-1}{\Upsilon+1}}}\ .
\en
The function ${\bar\phi}\paren{r_+,r_-}={\bar\phi}\paren{R\to l,r_+,r_-}$ is the electrostatic potential of the shell for $R=l$. The last term on Eq. \eqref{eq:PhiBI} is just the electrostatic potential of a shell of radius $R$, which coincides with Eq.
\eqref{eq:phiBIBTZ} when $r=R$. Thus, the thermodynamic electrostatic potential can be thought of as the blueshifted difference between the electrostatic potential of a shell of infinite radius and a shell of radius $R$.

Once the equations of state for the inverse temperature, the pressure and the thermodynamic electrostatic potential are known, it should be possible to obtain the entropy differential by ensuring that it is exact. Thus, in principle it should be possible to express the entropy differential as in Eq. \eqref{eq:dSrprm}. Notwithstanding, in order to do so, one must also know the functional relation between $\paren{M,Q}$ and $\paren{r_+,r_-}$. Given the form of the BI-BTZ metric in Eq. \eqref{eq:fBIBTZ}, such a task is far from being trivial and would require a numerical implementation -- which is outside the scope of the present work. However, given the fact that an analytical solution for the horizon radius of an extremal BI-BTZ black hole is known (see Eq. \eqref{eq:rexBIBTZ}), the analysis of the thermodynamics of an extremal shell should be feasible and shall be implemented next.

\subsubsection{Extremal Born-Infeld charged shell}
In the case of the extremal BI shell we have that the entropy differential is given by 
\be
\dd S=\beta_T\left[\paren{\frac{\p M_\ex}{\p r_\ex}}_R-\Phi_\ex\paren{\frac{\p Q_\ex}{\p r_\ex}}_R\right]\dd r_\ex\ ,
\en
where all the quantities are fixed to be evaluated at the extremal configuration. From direct calculation we obtain that the relevant parameters are
\begin{align}
Q_\ex&=\frac{r_\ex\sqrt{1+4 l^2\beta^2}}{2l^2\beta}\ , \label{eq:Qexrex}\\
m_\ex&=\frac{r_\ex^2}{4\beta^2l^4}\paren{1+4\beta^2l^2}\left[1-2\ln\paren{\frac{r_\ex}{4l^3\beta^2}\paren{1+4\beta^2l^2}}\right]\ , \label{eq:mexrex}\\
M_\ex&=\frac{R}{4 l}\paren{1-\sqrt{\left(2 \beta ^2 l^2\paren{1-\Upsilon_\ex}+1\right)+2 \frac{r_\ex^2}{R^2}\left( 1 +\frac{1}{4\beta ^2l^2}\right) \ln \left(2\beta^2l^2\frac{R}{r_\ex}\paren{\Upsilon_\ex-1}\right)}} \label{eq:Mexrex}
\end{align}
where $\Upsilon_\ex\equiv\Upsilon\paren{Q_\ex}$ as defined in Eq. \eqref{eq:Gammadef}, i.e., 
\be
\Upsilon_\ex=\sqrt{1+\frac{r_\ex^2}{R^2}\frac{\paren{1+4\beta^2 l^2}}{4\beta^4l^4}}\ .
\en
Given the above relations we obtain that the entropy differential is given by
\be \label{eq:dSexBI}
\dd S=
\frac{\sqrt{4\beta^2l^2+1}}{2\beta l^2}{\rm a}\paren{r_\ex}\paren{\frac{r_\ex}{4\beta l^2}\sqrt{4\beta^2l^2+1}-{\bar \phi}\paren{r_\ex}}\dd r_\ex\ .
\en
This relation is only dependent on $r_\ex$ and hence can be written as $\dd S=s\paren{r_\ex}\dd r_\ex$. In this way, we can think of the r.h.s. of Eq. \eqref{eq:dSexBI} as an \emph{entropy density}. Therefore, we expect for the entropy of an extremal BI shell to be dependent solely on $r_\ex$. Following this line of reasoning, we shall propose \emph{ans\"{a}tze} for the inverse temperature and the electrostatic potential that fulfill these requirements. 

One may obtain a lower bound on the electrostatic potential by considering that, in general, the inverse temperature is non-negative, i.e., ${\rm a}\paren{r_\ex}\ge0$, and a meaningful entropy differential must be non-negative. Hence, we directly obtain that the relation
\be \label{eq:boundphiBI}
{\bar \phi}\paren{r_\ex}\le\frac{r_\ex}{4\beta l^2}\sqrt{4\beta^2l^2+1}=\frac{Q
_\ex}{2}
\en
must hold, where we have used $r_\ex=2Q\beta l^2/\sqrt{4\beta^2l^2+1}$.

\begin{figure}[ht]
\includegraphics[width=0.87\linewidth]{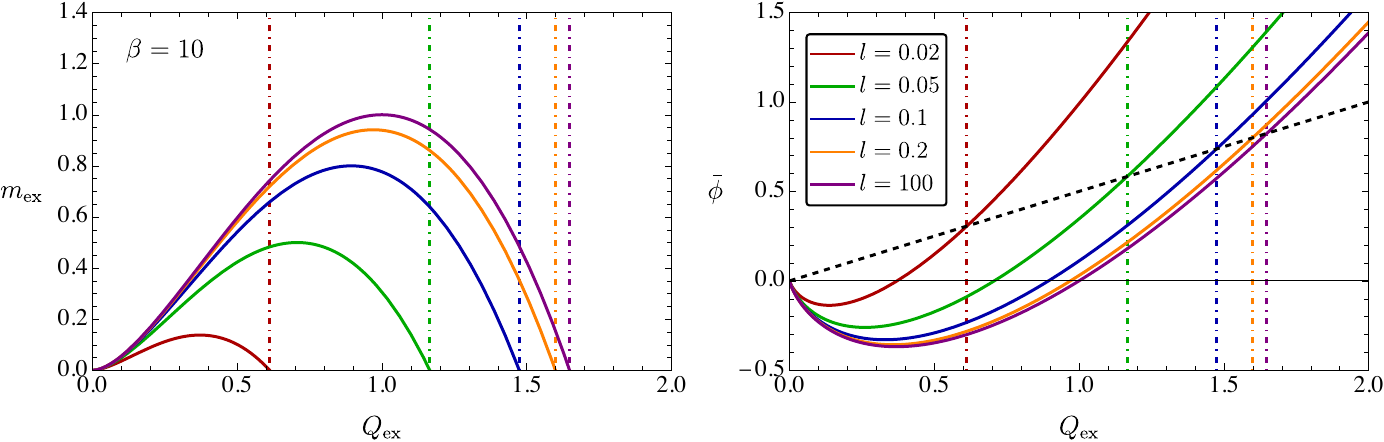}
\caption{(Left panel) Plot of the mass for the extremal configuration $m_\ex$ as a function of the charge for different values of $l$. Notice that there is a critical value in which $m_\ex=0$ for non-zero $Q$. (Right panel) Plot of the \emph{ansatz} for the electrostatic potential in Eq. \eqref{eq:ansatzphiBI} for different configurations of $l$. The dashed black line corresponds to the upper bound for the potential given by $Q/2$ (See Eq. \eqref{eq:boundphiBI}). In both plots we have fixed $\beta=10$ and the dotted vertical lines represent the points where $m_\ex=0$.}
\label{fig:prowBI}
\end{figure}
Since we are dealing with an extremal thin shell, we can no longer assume a Hawking-type equation of state for the temperature, because as $r_+\to r_-$ such an inverse temperature diverges. Hence, in order to obtain a non-divergent expression for the inverse temperature, let us consider an \emph{ansatz} of the form
\be \label{eq:ansatzaBI}
{\rm a}\paren{r_\ex}=\gamma \paren{m_\ex}^\delta\ ,
\en
where $\gamma$ is a parameter determined by the matter composing the shell and $\delta$ is a free parameter associated to the power law for the inverse temperature we are considering. At first instance, we are prompted to consider cases for $\delta>-1$ in order to avoid a singular behavior for the inverse temperature.
Another natural bound that takes place in the temperature chosen in Eq. \eqref{eq:ansatzaBI} is that $m_\ex\ge0$ in order to have always a non-negative inverse temperature. 

The next step is to fix an \emph{ansatz} for the electrostatic potential, ${\bar \phi}$. Eq. \eqref{eq:boundphiBI} sets an upper bound that the electrostatic potential must respect in order for the entropy density to be non-negative. Hence we can choose an \emph{ansatz} of the form
\be\label{eq:ansatzphiBI}
{\bar \phi}=Q \ln \left(\frac{Q \sqrt{4 \beta ^2 l^2+1}}{2 \beta  l}\right)\ .
\en
This particular choice for the electrostatic potential conforms to the bound given by Eq. \eqref{eq:boundphiBI}. Figure \ref{fig:prowBI} displays the behavior of the extremal mass, $m_\ex$ as a function of the charge for different values of $l$. The dashed vertical lines represent the values where $m_\ex=0$ for each configuration, hence the possible inverse temperatures are bounded to lie below this point. The right panel of Fig. \ref{fig:prowBI} displays the electrostatic potential chosen in Eq. \eqref{eq:ansatzphiBI}. Condition \eqref{eq:boundphiBI} gives us a maximum value that the electrostatic potential can take which is exactly attained at the intersection between each of the curves for the electrostatic potential and the vertical lines corresponding to $m_\ex=0$.

Substituting the \emph{ans\"{a}tze} for the inverse temperature and the electrostatic potential leads to an entropy density of the form
\begin{align} 
s(r_\ex)=&-\gamma\frac{r_\ex}{\beta^2}4^{-\delta -2} l^{-4 (\delta +1)} \left(1+4 \beta ^2 l^2\right)^{\delta +1} \left(\frac{r_\ex}{\beta }\right)^{2 \delta } \left[\ln \left(1+4 \beta ^2 l^2\right)+4 \ln \left(\frac{r_\ex}{4l^3\beta^2}\left(1+4 l^2\beta^2\right)\right)\right]\nonumber \\
&\times \left[1-2 \ln \left(\frac{r_\ex}{4l^3\beta^2}\left(1+4 l^2\beta^2\right)\right)\right]^{\delta }\ , \label{eq:sBIex}
\end{align}
that can be integrated in order to obtain the entropy of the shell as
\be \label{eq:Swb}
S=-2^{2 \delta -1}\gamma e^{\delta +1} (\delta +1)^{-\delta -2} \left(\frac{w^2}{\xi}\right)^{\delta +1} \left((\delta +1) \left(\ln \left(\xi\right)+2\right) \Gamma \left[\delta +1,\zeta\right]-2 \Gamma \left[\delta +2,\zeta\right]\right)+S_0\ ,
\en
where $\Gamma[a,z]$ is the incomplete Gamma function and we have defined the dimensionless parameters 
\be \label{eq:defwrl}
w\equiv l\beta\ , \quad \text{and}\quad r_l\equiv r_\ex/l\ ,
\en
and, for shortage
\begin{align}
\xi&\equiv1+4w^2\ , \\
\zeta&\equiv(1+\delta)\paren{1-2\ln\left[\frac{r_l\xi}{4w^2}\right]}\ ,
\end{align}

The integration constant $S_0$ is such that in the absence of matter, i.e., as $m\to0$, we obtain a null entropy. Thus, we obtain that the integration constant is $S_0=0$.

\begin{figure}[t]
\includegraphics[width=0.47\linewidth]{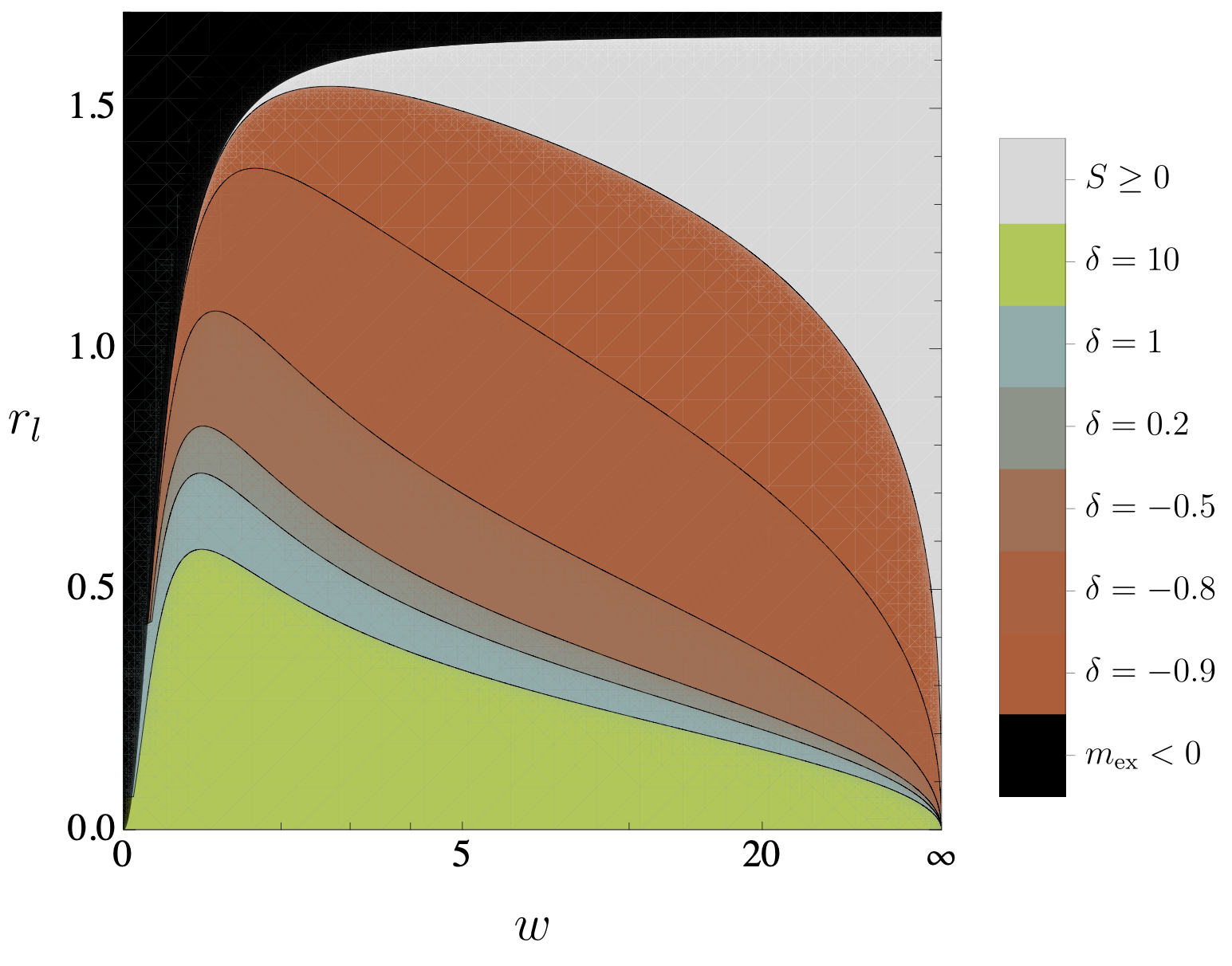}
\caption{Configurations for the extremal BI-BTZ shell in the $\paren{w,r_l}$ plane in which the entropy in Eq. \eqref{eq:Swb} is non-negative (colored regions), negative (gray region) and the ADM mass is negative (black region). Notice that as $\delta$ grows, the allowable configurations leading a non-negative entropy, decrease.}
\label{fig:plotSpos}
\end{figure}

Notice that the latter entropy needs to be non-negative, and such a requirement shall set restrictions on the possible values the parameters can take. 
Given the above expression, it is possible to study the configurations in the $\paren{w,r_l}$ plane that allow for a positive entropy. 
In Fig. \ref{fig:plotSpos}, we display the regions where $S\ge0$ for diverse values of $\delta$ in the $\paren{w,r_l}$ plane. Notice that this plane can allow for regions where $S<0$ (which are displayed as the gray regions) and where $m_\ex<0$ (displayed as the black region), that shall be discarded when analyzing the thermodynamical stability. It can be seen that, as $\delta$ grows, the allowable domain of non-negative entropy decreases. In the Figure, the limiting curves for the colored regions are the contours where $S=0$. Let us note that as $w\to\infty$, i.e., as the Maxwell limit is recovered, all $S=0$ curves coincide, indicating that in the Maxwell limit, the only allowable entropy function for an extremal shell is that of $S=0$ at $r_l=0$, regardless of the value of $\delta$. 
\section{Thermodynamical stability} \label{sec:thermstab}
Considering that the entropy of the shell is characterized by $S=S\paren{M,Q,P}$, it is possible to obtain conditions for exchanges of $M$, $Q$ and $P$ that ensure the local stability for the thermodynamic system. These conditions are related to demanding that the resulting entropy after the exchange of some matter is less than the entropy before the exchange and, therefore, to the concavity of the entropy function. Adopting the formalism developed by Callen (cf. \cite{Callen1985}), we have that the thermodynamical stability conditions for such a system are given by the inequalities
\begin{align}
\paren{\frac{\p^2 S}{\p M^2}}_{P,Q}\le0\ ,\\
\paren{\frac{\p^2 S}{\p P^2}}_{M,Q}\le0\ , \\
\paren{\frac{\p^2 S}{\p Q^2}}_{M,P}\le0\ , \\
\paren{\frac{\p^2 S}{\p M^2}}\paren{\frac{\p^2 S}{\p P^2}}-\paren{\frac{\p^2 S}{\p M\p P}}^2\ge0\ , \\
\paren{\frac{\p^2 S}{\p P^2}}\paren{\frac{\p^2 S}{\p Q^2}}-\paren{\frac{\p^2 S}{\p P\p Q}}^2\ge0\ , \\
\paren{\frac{\p^2 S}{\p M^2}}\paren{\frac{\p^2 S}{\p Q^2}}-\paren{\frac{\p^2 S}{\p M\p Q}}^2\ge0\ , \\
\paren{\frac{\p^2 S}{\p M^2}}\paren{\frac{\p^2 S}{\p P\p Q}}-\paren{\frac{\p^2 S}{\p M\p P}}\paren{\frac{\p^2 S}{\p M \p Q}}\ge0\ .
\end{align}
Given the fact that the material mass of the shell is related to the linear density by Eq. \eqref{eq:lambda}, and the shell's charge is related to the linear charge density by Eq. \eqref{eq:lambdae}. In principle, these densities are not related and shall be treated independently, which of course leads to considering the above stability conditions independently. However, for computational matters, we can assume a linear relation between $\lambda$ and $\lambda_e$ such that $\lambda_e=\alpha \lambda$, which in a way gives us the charge-to-mass ratio for the shell. Thus, the resulting stability conditions using this parametrization are \cite{Reyes2022,Eiroa2024}
\begin{align}
S_{PP}\le0\ ,& \label{eq:Thstab1}\\
\alpha^2 S_{QQ}+2\alpha S_{MQ}+S_{MM}\le0\ , & \label{eq:Thstab2}\\
S_{PP}\paren{S_{MM}+\alpha^2S_{QQ}+2\alpha S_{MQ}}-\paren{\alpha S_{PQ}+S_{PM}}^2\ge0\ .& \label{eq:Thstab3}
\end{align}

\subsection{Maxwell charged thin shell}

Considering an inverse temperature for the shell of the Hawking type as in Eq. \eqref{eq:aHawking} leads to an entropy solely dependent on the shell's gravitational radius,
\be
S=\gamma r_+\ ,
\en
where in turn $r_+=r_+\paren{R,M,Q}$ and the thermodynamical stability conditions in Eqs. \eqref{eq:Thstab1}-\eqref{eq:Thstab3} can be readily evaluated. Straightforward computation shows that the following relations hold 
\begin{align}
\frac{R^2}{M^2}S_{PP}&=\paren{\alpha^2S_{QQ}+2\alpha S_{MQ}+S_{MM}}\ , \\
\frac{R}{M}S_{PP}&=-\paren{\alpha S_{PQ}+S_{PM}}\ .
\end{align}
Using these expressions, condition \eqref{eq:Thstab3} is trivially satisfied and the only relevant condition for determining the full thermodynamical stability reduces to
\be\label{eq:SPPa}
S_{PP}\le0\ , 
\en
which in turn can be written as
\be
\frac{\Omega^2 \left[\alpha ^4 (1-2 \kappa )+32 \alpha ^2 \kappa ^2+\alpha ^4 \Omega^2 (\kappa -1)^2+\Omega \left(\alpha ^4 \left(3 \kappa ^2-4 \kappa +2\right)-32 \alpha ^2 \kappa ^2+256 \kappa ^2\right)-256 \kappa ^2\right]}{2 (\Omega+1)^3 (\kappa -1) \left(\alpha ^2-16 \kappa \right) \left(-\alpha ^2 \Omega\right)^{3/2}}\le0\ .
\en
where we have defined $\Omega\equiv W_{-1}\left(-\exp\paren{\frac{2 \left(\alpha ^2-8 (\kappa +1)\right)}{\alpha ^2 (\kappa -1)}} \left(\alpha ^2-16 \kappa \right)^2/\paren{16 \alpha ^2 (\kappa -1)^2}\right)$. Considering that the matter respects the WEC we have that the latter condition is only satisfied for 
\be
\alpha>4\ ,\  \kappa<1\ ,\ \text{and}\ r_+> r_\ex\ ,
\en
so that all configurations for $\alpha>4$ and above the critical line that determines the (in)existence of horizons are thermodynamically stable.

\begin{figure}[ht]
\includegraphics[width=0.47\linewidth]{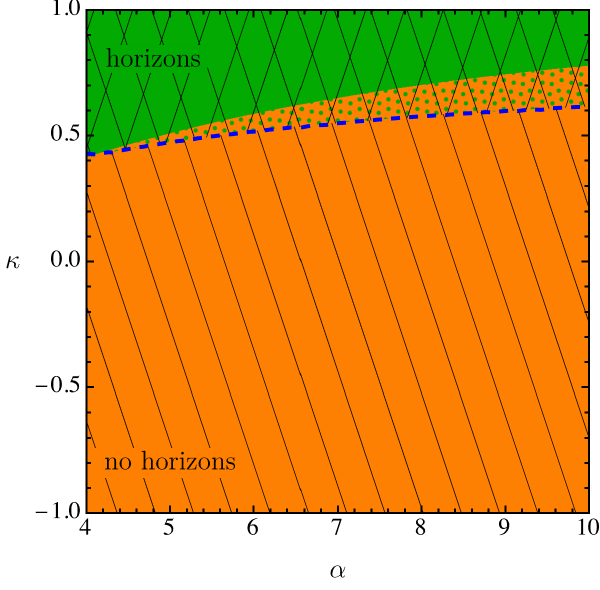}
\caption{Regions of thermodynamical and dynamical stability for the RN thin-shell with a Hawking-type entropy (green and orange regions, respectively). These conditions lead to configurations in the parameter space which correspond to the fulfillment of the two stability criteria (orange dotted regions), that can be regarded as more stable than those in which only one stability criteria holds.}
\label{fig:pRNTDS}
\end{figure}

Figure \ref{fig:pRNTDS} displays the regions of thermodynamical stability joined with those of dynamical stability. It can be seen that this choice of entropy is fully thermodynamically stable where the entropy is meaningful (i.e., where there are horizons). Furthermore, since part of the dynamical stability region lies over configurations with horizons, there is a region for the parameters that allow for the existence of both stability criteria, and hence can be regarded as more stable with respect to solely demanding one type of stability.

\subsection{Extremal Born-Infeld charged thin shell}
Let us consider the case of thermodynamical stability for the extremal BI shell. Notice that a Hawking type inverse temperature cannot be considered since for an extremal BH, such a function must diverge, leading to an impossibility for obtaining the corresponding entropy from thermodynamical considerations. Then, taking into account an inverse temperature of the power law type as in Eq. \eqref{eq:ansatzaBI} and a corresponding \emph{ansatz} for the electrostatic potential of the form \eqref{eq:ansatzphiBI} leads to the entropy in Eq. \eqref{eq:Swb}; that has been conveniently rewritten in terms of two dimensionless parameters $w\equiv l\beta$ and $r_l\equiv r_\ex/l$ such that it is possible to explore the domain in the $\paren{w,r_l}$ plane where the corresponding entropy is non-negative, as displayed in Fig. \ref{fig:plotSpos}. 

In view of the fact that the entropy for the extremal shell, as displayed in Eq. \eqref{eq:Swb}, is a function of the extremal radius alone, it becomes necessary to write this entropy as a function of the relevant extensive parameters in order to evaluate the thermodynamical stability conditions. We can start considering that, from Eqs. \eqref{eq:Qexrex} and \eqref{eq:mexrex}, the extremal gravitational radius can be written in terms of the extremal mass and charge as
\be
r_\ex=\frac{4l^3\beta^2}{1+4\beta^2l^2}\exp\left[\paren{1-m_\ex/Q_\ex^2}/2\right]\ .
\en
Additionally, in order to compute the relevant derivatives to evaluate the thermodynamical stability conditions, it is necessary to write the entropy as a function of  $\paren{M_\ex,Q_\ex,R}$. This can be achieved by writing $m_\ex$ in the same fashion as in Eq. \eqref{eq:mMRQ} as
\be \label{eq:mMRQex}
m_\ex=8\frac{M_\ex R}{l}-16M_\ex^2+2R^2\beta^2\paren{1-\Upsilon_\ex}+Q_\ex^2\paren{1-2\ln\left[\frac{R}{2l}\paren{1+\Upsilon_\ex}\right]}\ ,
\en
such that the entropy in Eq. \eqref{eq:Swb} is now a function of $\paren{M_\ex,Q_\ex;R}$ and the thermodynamical stability conditions can be directly calculated.

Up to now, the dynamical stability conditions have been reduced to sets of configurations of three parameters, namely $\paren{\kappa,\alpha,z}$, that are related to the equation of state, the charge-to-mass ratio and the departures from Maxwell's theory, respectively. On the other hand, the entropy is a function of $r_\ex$ which in turn is a function of $\paren{M_\ex,Q_\ex;R}$, as discussed above. Once the corresponding derivatives for evaluating the thermodynamical stability conditions are calculated, it is instructive to write the respective relations in terms of the dimensionless parameters $\paren{w,r_l,\sigma}$, where
\be
\sigma\equiv R\beta\ ,
\en 
and $w$ and $r_l$ have been introduced in Eq. \eqref{eq:defwrl}. In principle there is no relationship among these two sets of parameters and, so far, the stability criteria cannot be compared. However, let us note that the junction conditions allow us to parametrize the charge-to-mass ratio as $Q_\ex=\alpha M_\ex$, so that it is possible to obtain an expression for $\alpha$ in terms of $\paren{w,r_l,\sigma}$. Furthermore, the lower bound on $\alpha$, given by Eq. \eqref{eq:condalphaBI} and the coefficient $\kappa$ appearing in the equation of state can also be written in terms of the new set of parameters and hence both the dynamical and thermodynamical stability criteria can be evaluated for distinct configurations of $\paren{w,r_l,\sigma}$. Noticing that the parameter $\sigma$ is bounded from below as (see Appendix \ref{sec:ApA} for details). 
\be \label{eq:boundsigma}
\sigma>r_l w\ .
\en

\begin{figure}[t]
\includegraphics[width=0.87\linewidth]{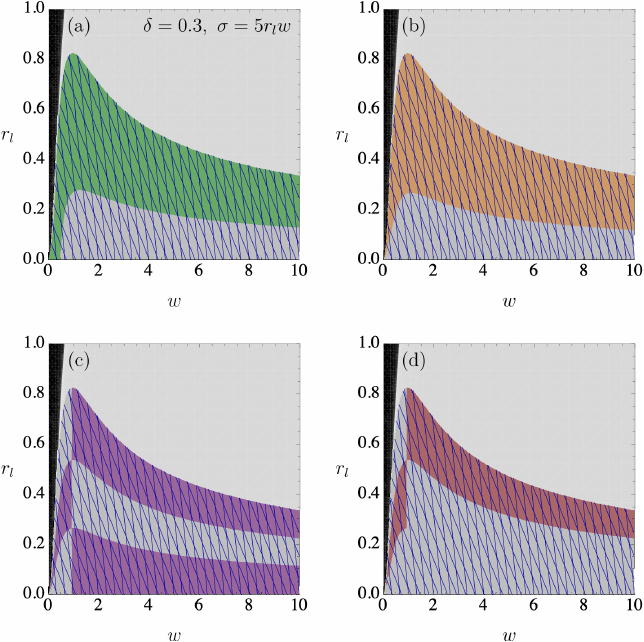}
\caption{Regions of thermodynamical stability for the conditions (a) $S_{PP}\le0$, (b) $\alpha^2 S_{QQ}+2\alpha S_{MQ}+S_{MM}\le0$, (c) $S_{PP}\paren{S_{MM}+\alpha^2S_{QQ}+2\alpha S_{MQ}}-\paren{\alpha S_{PQ}+S_{PM}}^2\ge0$ and (d) the three conditions are satisfied. We have set $\delta=0.3$ and $\sigma=5 r_l w$. }
\label{fig:tstabBI}
\end{figure}

The physically relevant thermodynamical stability regions in the given parameter space come from matching conditions \eqref{eq:Thstab1} - \eqref{eq:Thstab3} with the region where $S\ge0$ which is given in Fig. \ref{fig:plotSpos}, alongside with the regions where $\alpha\paren{w,r_l,\sigma}>\alpha_\text{min}\paren{w,r_l,\sigma}$. Taking into account bound \eqref{eq:boundsigma}, it is possible to explore the thermodynamical stability in the $\paren{w,r_l}$ plane by considering values of $\sigma$ that conform to the latter bound. Each of the thermodynamical stability conditions must be evaluated independently and the corresponding configuration under scrutiny should be regarded as thermodynamically stable if all three conditions are simultaneously satisfied. Figure \ref{fig:tstabBI} displays the different thermodynamical stability conditions evaluated for a particular configuration of $\delta=0.3$ and $\sigma=5r_lw$. It can be seen that, for this particular choice of parameters, the conditions \eqref{eq:Thstab1} and \eqref{eq:Thstab2} are qualitatively similar. Thus, just as in the Maxwell case, considering exchanges in $R$ and in $\paren{M,Q}$ are equivalent. On the other hand, the consideration of exchanges in all the extensive parameters simultaneously leads to a different region for thermodynamical stability, which reduces the possible stable configurations for the full thermodynamical stability.

A completely stable configuration is regarded as one that fulfills both dynamical and thermodynamical stability conditions simultaneously. Once the thermodynamical stability for the extremal BI thin shell is given in terms of the parameters $\paren{w,r_l,\sigma}$, and the dynamical stability for the linear equation of state was established in terms of the $\paren{\alpha,\kappa,z}$ parameters, it becomes necessary to obtain a map between these. In Appendix \ref{sec:ApA} such a map is developed and, within the new set of parameters $\paren{w,r_l,\sigma}$, all allowable configurations for extremal BI thin shells are dynamically stable.

\begin{figure}[t]
\includegraphics[width=0.87\linewidth]{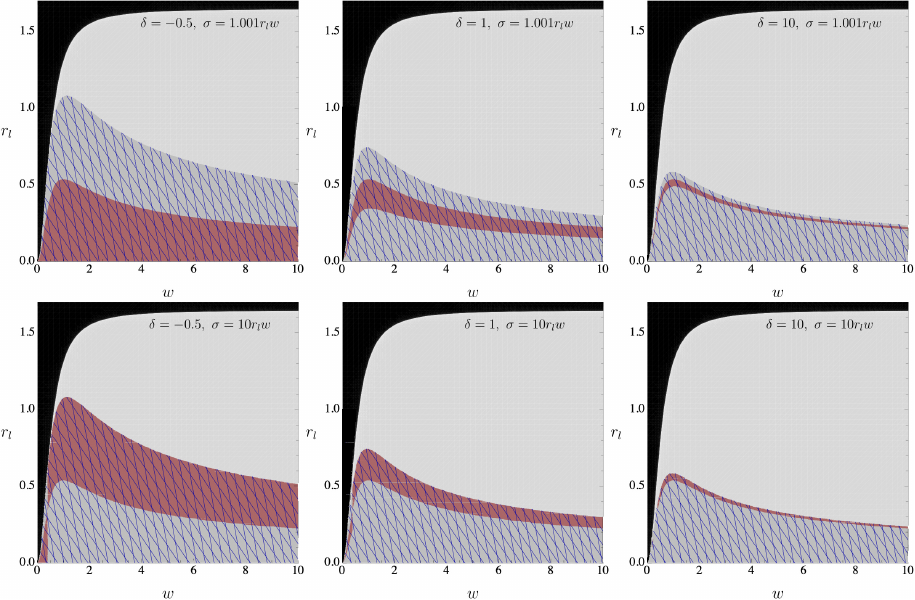}
\caption{ Regions of thermodynamical stability in the $\paren{w,r_l}$ plane for distinct configurations of $\paren{\delta,\sigma}$. The meshed region represents non-negative entropy, while the black region depicts negative mass configurations. As $\delta$ grows, the stability regions shrink, while the overall effect of increasing $\sigma$ is a displacement of the stabilty regions. }
\label{fig:gplotstab}
\end{figure}

Consequently, the completely stable configurations will be those intersections between the whole parameter space (which belongs to the dynamically stable configurations) and the thermodynamically stable configurations. Then, it becomes necessary to address the parametric dependence of the thermodynamically stable configurations within this parametrization. In Fig. \ref{fig:gplotstab}, we display the thermodynamically stable configurations in the $\paren{w,r_l}$ plane, for different values of $\paren{\delta,\sigma}$. It can be seen that small values of $\delta$ correspond to larger regions in the parameter space that are stable, while, as $\delta$ grows this region shrinks. On the other hand, the variation in $\sigma$, respecting the bound \eqref{eq:boundsigma}, corresponds to a displacement in the $r_l$ direction for the whole stability region. Such displacement stops once $\sigma$ has attained a sufficiently high value. 

\section{Conclusions and perspectives}
\label{sec:conclusions}

In the present work, we have studied the dynamical and thermodynamical stability conditions for a $\paren{2+1}$-dimensional charged thin shell within BI electrodynamics. In the process of constructing the thin shell spacetime, we have re-derived the BI-BTZ black hole line element ensuring that the corresponding Maxwell-BTZ and static BTZ black holes are obtained in the $\beta\to\infty$ and $Q\to0$ limits, respectively.

We have derived the conditions for dynamical stability for the BI shell composed of matter obeying a linear equation of state. This type of equation of state allows us to establish a set of parameters, $\paren{\alpha,\kappa,z}$ (which has been previously introduced in the context of thin shell wormholes within Maxwell electrodynamics in \cite{Eiroa2024}), where $z$ is related to the nonlinear departures from Maxwell's theory. Within this new parametrization, we have obtained similar results as in Ref. \cite{Eiroa2013} for this type of stability. Regarding to the cases we address in the present work, the dynamical stability for the Maxwell charged thin shell (see Fig. \ref{fig:RNBTZdynstab}), all the region of parameters where the thin shell is composed by an outer manifold lacking of any horizons is dynamically stable, while a part of the region that corresponds to the outer manifold consisting of a spacetime with horizons is also stable. As BI nonlinearities are taken into account (see Fig. \ref{fig:BIBTZdynstab}), the dynamical stability region diminishes in the domain where the outer manifold has horizons, while the region consisting of no horizons is still completely stable. As a matter of fact, as $z$ grows the maximum allowable value of $\kappa$ for having a stable configurations tends to 0, i.e., only the regions with negative pressures are dynamically stable. Let us also note that the critical lines defining the extremal thin shell vary as the nonlinearity parameter $z$ varies. In addition, the extremal lines are always contained within the dynamically stable region, independent of the value of $z$. This last feature is better displayed in the $\paren{w,r_l,\sigma}$ parametrization.

Furthermore, we have developed the thermodynamics and studied the thermodynamical stability for the charged thin shells in both Maxwell and BI electrodynamics. In the particular case of Maxwell electrodynamics, using an equation of state for the inverse temperature of the Hawking type, we have been able to derive an entropy for the shell that is coincident with that of the charged BH in $\paren{2+1}$ dimensions. This particular choice for the inverse temperature allows us to study configurations that are thermodynamically stable in the regime where the outer manifold encloses horizons and $R>r_+$. Figure \ref{fig:pRNTDS} displays both thermodynamically and dynamically stable configurations and it can be seen that there is a region in the parameter space that is completely stable. To the best of our knowledge, the study of complete stability for a charged thin shells in $\paren{2+1}$ dimensions constitutes novel results.

We have also developed the thermodynamics of a generic BI thin shell. However, due to the lack of an analytic solution for the relation between the gravitational radii and the mass and charge parameters, it is no longer possible to obtain straightforward relations for the temperature and electrostatic potential equations of state and such a task would require further numerical work. Notwithstanding, due to the fact that the extremal BI thin shell has a well known relation for the gravitational radius in terms of the mass and charge, we have obtained that the shell's entropy is necessarily a function of $r_\ex$ alone, coinciding with the treatment of other kinds of extremal shells in other dimensions (cf. \cite{Lemos2013}). Furthermore, we have been able to propose some well suited \emph{ans\"{a}tze} that conforms to the physically realistic scenario that the \emph{entropy density} is non-negative. Furthermore, using a power law equation of state for the inverse temperature, we have been able to find a closed analytic solution for the extremal BI shell entropy. Thus, we have also been able to study the thermodynamical stability configurations for such a shell.

For the extremal BI thin shell, we have obtained that the thermodynamically stable regions are always contained in the dynamically stable regions. As well, these thermodynamically stable regions are relevant only when contained inside the domain where the entropy is non-negative. Thus, for the extremal BI shells, we find that the thermodynamically stable regions have a well defined parametric dependence: namely, as $\delta$ grows the region shrinks and as $\sigma$ grows the region is displaced upwards in the $\paren{w,r_l}$ plane. Finally, the complete stability for the extremal BI shell is better presented in the $\paren{w,r_l,\sigma}$ parametrization, which is better suited for studying the dynamical stability of the extremal shell than the $\paren{\alpha,\kappa,z}$ parametrization since we have a single extremal curve for each $z$, while all extremal curves are contained in the $\paren{w,r_l}$ plane. Thus, in a way, this new parametrization condenses the information for the dynamical stability of extremal shells in such a way that the complete stability for them can be easily studied.

Despite the apparent simplicity of studying systems in $\paren{2+1}$-dimensional spacetimes, the thermodynamics for a BI thin shell cannot be studied analytically in the most general regime. Surprisingly enough, the particular case of an extremal BI shell can be analytically studied. As future avenues of research, let us note that the entropy function for the extremal BI shell recovers $S=0$ for the Maxwell limit; this particular scenario is in accordance with numerous claims that the entropy of an extremal black hole is zero (cf. \cite{Lemos2015a}) and deserves further attention. Also, we intend to generalize our results to generic BI shells in $\paren{2+1}$ dimensions and analyze the complete stability for a BI thin shell in $\paren{3+1}$ dimensions in future work.

\begin{acknowledgments}
	This work was supported by the National Council for Scientific and Technological Development - CNPq and FAPERJ - Fundaçāo Carlos Chagas Filho de Amparo à Pesquisa do Estado do Rio de Janeiro, Processo SEI 260003/014960/2023 (MLP). MLP thanks CINVESTAV-IPN (Mexico City) for their hospitality, where part of this work was written.
\end{acknowledgments}
\appendix


\section{Mapping the parameter space $\paren{\alpha,\kappa,z}$ to $\paren{w,r_l,\sigma}$ and reformulating dynamical stability for the extremal Born-Infeld thin-shell} \label{sec:ApA}
The dynamical stability regions for the BI thin-shell are obtained by demanding that $V''\paren{R_0}>0$ in Eq. \eqref{eq:VppcondBI}. Upon assuming that the matter obeys a linear equation of state of the form $p=\kappa \lambda$ and that the shell's mass, $M$, and charge, $Q$, can be related by a linear parameter as $Q=\alpha M$, the corresponding condition can be written in terms of a function dependent on three dimensionless parameters, $\paren{\alpha,\kappa,z}$, where $z\equiv\lambda_0/\beta$. The dynamical stability regions for variable $z$ are displayed in Fig. \ref{fig:BIBTZdynstab}.

On the other side, the thermodynamical stability regions are obtained through conditions \eqref{eq:Thstab1} - \eqref{eq:Thstab3} along with the requirement that the entropy function of the shell is non-negative for the stable configurations. As displayed in Fig. \ref{fig:gplotstab}, these regions are better characterized within the parameters $\paren{w,r_l,\sigma}$, which are defined as 
\be
w\equiv l\beta\ ,\ r_l\equiv r_\ex/l\ \text{and}\ \sigma\equiv R\beta\ .
\en
Ever since the sets of parameters $\paren{\alpha,\kappa,z}$ and $\paren{w,r_l,\sigma}$ are not obviously related, in this Appendix we display how these two sets are mapped into each other such that the dynamical stability regions can be displayed in the thermodynamically adequate parameters $\paren{w,r_l,\sigma}$.

Assuming a linear relation among the material mass and charge, for the the extremal BI thin-shell we have $M_\ex=2\pi R \lambda_\ex$ and $Q_\ex=2\pi R\alpha \lambda_\ex$, where both quantities, given in Eqs. \eqref{eq:Mexrex} and \eqref{eq:Qexrex}, can be expressed in terms of the parameters $\paren{w,r_l,\sigma}$. Thus, given these expressions, it is possible to solve for $\alpha$ in terms of the new parameters, as
\be \label{eq:alphawl}
\alpha=\frac{4 r_l \sqrt{4 w^2+1}}{2 \sigma -\sqrt{4 \sigma  \left(-w \sqrt{4 r_l^2 w^2+r_l^2+4 \sigma ^2 w^2}+\sigma +2 \sigma  w^2\right)+2 r_l^2 \left(4 w^2+1\right) \ln\left(\frac{\sqrt{4 r_l^2 w^2+r_l^2+4 \sigma ^2 w^2}-2 \sigma  w}{r_l}\right)}}\ ,
\en
where we have defined $\sigma\equiv R\beta$. Recalling that $\alpha$ has a lower bound given by Eq. \eqref{eq:condalphaBI}, which, for completeness is 
\be
\alpha>\alpha_\text{min}=\sqrt{16+256\pi^2 z^2}\ ,
\en
we directly obtain that, from the definition $z=\lambda_0/\beta$, in the extremal case 
\be \label{eq:zwl}
z=z\paren{w,r_l,\sigma}=M_\ex/\paren{2\pi \sigma}\ , 
\en
 and the lower bound $\alpha_\text{min}$ is
\be
\alpha_\text{min}=\sqrt{\frac{\left(\sqrt{4 \sigma  \left(\sigma  \left(2 w^2+1\right)-w \sqrt{4 w^2 \left(r_l^2+\sigma ^2\right)+r_l^2}\right)+2 r_l^2 \left(4 w^2+1\right) \ln \left(\frac{\sqrt{4 w^2 \left(r_l^2+\sigma ^2\right)+r_l^2}-2 \sigma  w}{r_l}\right)}-2 \sigma \right)^2}{\sigma ^2 w^2}+16}\ .
\en
Finally, since we are considering an equation of state of the form $p=\kappa \lambda$, for the extremal shell we have 
\begin{align}
\kappa&=p_\ex/\lambda_\ex \nonumber\\
&=-\frac{\left(\frac{2 \sigma ^2 w (\varepsilon+2 \sigma  w)}{r_l^2}+\left(4 w^2+1\right) (\sigma -\varepsilon w)\right) g(w,r_l,\sigma )+\left(4 w^2+1\right) \left(2 \sigma  w (\varepsilon+2 \sigma  w)+r_l^2 \left(4 w^2+1\right)\right) \ln \left(\frac{r_l}{d-2 \sigma  w}\right)}{\frac{\varepsilon (\varepsilon+2 \sigma  w)  \left(r_l^2 \left(4 w^2+1\right) \ln \left(\frac{\varepsilon-2 \sigma  w}{r_l}\right)+2 \sigma  \left(\sigma  \left(2 w^2+1\right)-\varepsilon w\right)\right)}{2 r_l^2 \sigma }g(w,r_l,\sigma )}\ , \label{eq:kappawrl}
\end{align}
where we have defined 
\begin{align}
\varepsilon&=\varepsilon\paren{w,r_l,\sigma}\equiv\sqrt{4 w^2 \left(r_l^2+\sigma ^2\right)+r_l^2}\ , \\
g\paren{w,r_l.\sigma}&\equiv\sqrt{2 r_l^2 \left(4 w^2+1\right) \ln \left(\frac{\varepsilon-2 \sigma  w}{r_l}\right)+4 \sigma  \left(\sigma  \left(2 w^2+1\right)-\varepsilon w\right)}-2 \sigma\ .
\end{align}
Hence, relations \eqref{eq:alphawl}, \eqref{eq:zwl} and \eqref{eq:kappawrl} correspond to the map of the parameters $\paren{\alpha,\kappa,z}$ into $\paren{w,r_l,\sigma}$. 

Expressing the dynamical stability condition, Eq. \eqref{eq:VppcondBI}, in terms of the newly introduced parameters results in the complete fulfillment of the condition in all possible allowable regions. This region must be matched with that in which $\alpha>\alpha_\text{min}$, however, noticing that the construction of the shell demands that $r_\ex<R$, we have a lower bound on $\sigma$ given by 
\be
r_l=\frac{r_\ex}{l}<\frac{R}{l}=\frac{\sigma}{w}\Rightarrow \sigma>r_l w\ .
\en
Straightforward computation shows that even for values of $\sigma$ close to the minimal value given by the latter relation, the whole parameter space corresponds to configurations that fulfill the dynamical stability condition.

At last, let us note that this new parametrization allows for a complete study of the extremal thin-shell spacetimes for both dynamical and thermodynamical stabilities. In a way, this new parametrization condenses all the possible extremal shell curves (defined as the blue curves in Fig. \ref{fig:BIBTZdynstab}) into a single region. Thus, all possible extremal thin-shells are dynamically stable when considering a linear equation of state.
\bibliography{BIBTZ}
\end{document}